\documentclass{article}
\usepackage{ijcai24}

\usepackage{times}
\usepackage{soul}
\usepackage{url}
\usepackage[hidelinks]{hyperref}
\usepackage[utf8]{inputenc}
\usepackage[small]{caption}
\usepackage{graphicx}
\usepackage{amsmath}
\usepackage{amsthm}
\usepackage{booktabs}
\usepackage{algorithm}
\usepackage{algorithmic}
\usepackage[switch]{lineno}
\usepackage{float}
\usepackage{color}

\usepackage{amssymb}
\usepackage{comment}
\usepackage{multicol}
\usepackage[title]{appendix}

\usepackage{tabularray}
\usepackage{listings}

\title{Emergence of Social Norms in Generative Agent Societies:\\
Principles and Architecture}

\author{
Siyue Ren$^{1,\dag}$
\and
Zhiyao Cui$^{2,\dag}$\and
Ruiqi Song$^2$\and
Zhen Wang$^{1,2,3,*}$\And
Shuyue Hu$^{4,*}$
\affiliations
$^1$School of Mechanical Engineering, Northwestern Polytechnical University\\
$^2$School of Cybersecurity, Northwestern Polytechnical University\\
$^3$School of Artificial Intelligence, OPtics and ElectroNics(iOPEN), Northwestern Polytechnical University\\
$^4$Shanghai Artificial Intelligence Laboratory
\emails
\{rensiyue, zhiyao, songruiqi\}@mail.nwpu.edu.cn,
w-zhen@nwpu.edu.cn,
hushuyue@pjlab.org.cn
}

\usepackage{fancyhdr}
\usepackage{geometry}
\begin{document}

\maketitle

\begin{abstract} 
\renewcommand{\thefootnote}{\fnsymbol{footnote}}
\footnotetext[2]{Co-first, equal contributions} 
\footnotetext[1]{Corresponding authors}
    Social norms play a crucial role in guiding agents towards understanding and adhering to standards of behavior, thus reducing social conflicts within multi-agent systems (MASs). However, current LLM-based (or generative) MASs lack the capability to be normative. In this paper, we propose a novel architecture, named \textit{CRSEC},
    to empower the emergence of social norms within generative MASs.
    Our architecture 
    consists of four modules: Creation \& Representation, Spreading, Evaluation, and Compliance. 
    This addresses  
    several important aspects of the emergent processes all in one: 
    (i) where social norms come from, (ii) how they are formally represented, (iii) how they spread through agents' communications and observations, (iv) how they are examined with a sanity check and synthesized in the long term, and (v) how they are incorporated into agents' planning and actions.  
    Our experiments deployed in the Smallville sandbox game environment demonstrate the capability of our architecture to establish social norms and reduce social conflicts within 
 generative MASs.
The positive outcomes of our human evaluation, conducted with 30 evaluators, further affirm the effectiveness of our approach. Our project can be accessed via the following link: \href{https://github.com/sxswz213/CRSEC}{\color{blue}{https://github.com/sxswz213/CRSEC}}.
\end{abstract}

\section{Introduction}
In human societies, social norms, which are standards of behavior shared within a social group \cite{sherif1936psychology},  have shaped almost every aspect of our daily life, from the language we speak and the etiquette we drive to the amount we tip. Without social norms, people may feel confused about how to behave appropriately in social situations and consequently social conflicts may arise \cite{lewis1969convention}.
Over the past decades, the study of social norms has attracted much interest in a variety of disciplines, such as economics \cite{young2015evolution},  cognitive science \cite{hawkins2019emergence},  complex system science \cite{centola2018experimental}, and computer science \cite{morris2019norm}.
Across these studies, a central question is: how do social norms spontaneously emerge
from social interactions of humans or agents?

This paper studies the emergence of social norms within a generative multi-agent system (MAS), i.e. a system of agents that are powered by large-language models (LLMs). 
The deployment of MASs in real-world situations raises the need for these systems to be normative---the capability of empowering agents to understand certain standards of behavior and behave appropriately according to the standards \cite{boella2008introduction,criado2011open,oldenburg2024learning}. 
Imagine that agents within a system interact with other agents or humans to accomplish some tasks; for the system to be truly accepted and embraced by humans, such a capability will be crucial, as it can reduce conflicts within systems, enable more effective coordination among agents (potentially including humans), and allow humans to anticipate the system's behaviors---a key means to improve human trust in the system \cite{awad2018moral,ajmeri2020elessar,chugunova2022we,liullm}.

\begin{figure*}[htbp]
    \centering
\includegraphics[width=0.85\textwidth]{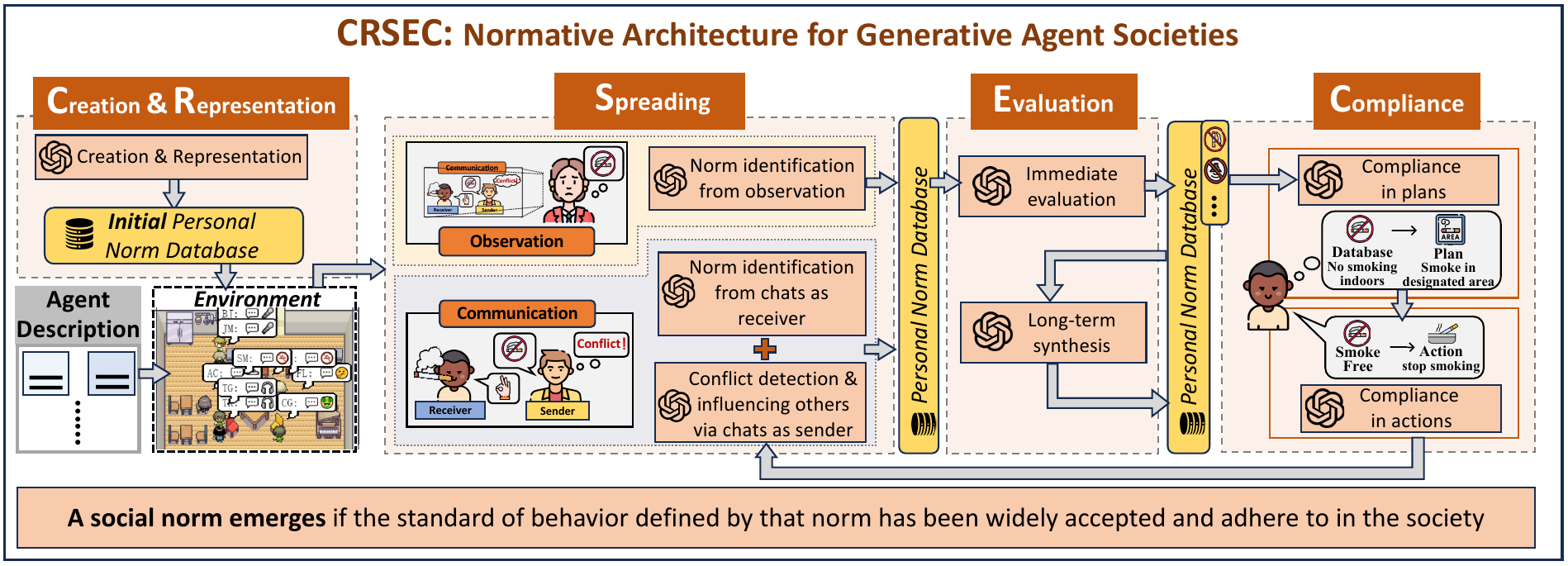}
    \vspace{-0.2cm}
    \caption{CRSEC: our architecture for the emergence of social norms in generative agent societies. Initially, by the \textit{Creation \& Representation} module, norm entrepreneurs create their personal norms and store them into their databases. By the \textit{Spreading} module, some agents proactively influence others to adopt their personal norms through initiating communication with others, while others can identify those norms from their chats and observations. The identified norms then undergo an immediate evaluation in the \textit{Evaluation} module.
The \textit{Compliance} module enables agents to generate plans and actions, with the norms bearing in mind.
The normative actions, in turn, can influence other agents' observations and thus reinforce the spreading of norms. In addition, from time to time, agents perform long-term synthesis to keep their personal norms compact and concise. }
    \vspace{-0.3cm}
    \label{fig 1: Normative architecture}
\end{figure*}

Since LLMs are trained on extensive corpora of human text, 
it is not surprising that they may inherently embed social norms \cite{schramowski2022large,guo2023data}. 
One might thus challenge the importance of fostering the emergence of social norms within generative MASs.
While LLMs can capture social norms, it has also been shown that LLMs do not adequately understand social norms, especially the culture-specific ones \cite{ramezani2023knowledge,hammerl2022multilingual}. This deficiency can provoke conflicts among generative agents, particularly when their base LLMs are trained on text corpora from diverse cultural backgrounds.
Moreover, as generative agents increasingly become more personalized (such as functioning as personal assistants) and represent humans in social situations, it is natural to expect that these agents, reflecting the values and preferences of their human users, will encounter social conflicts similar to those experienced by humans.
To tackle these challenges,
approaches must go beyond merely embedding LLMs with human norms or aligning them to such norms \cite{liu2024training,li2024agent}; rather, they should also be able to foster the emergence of social norms within generative MASs so that generative agents can establish their own standards of behavior out of their interactions and adhere to these standards to address those conflicts.

How can we empower generative MASs with the capability to foster social norm emergence? 
We argue that the key is to instigate an emergent process---generative agents, starting with initially only a few adopting certain standards of behavior, influence others and propagate these standards, ultimately resulting in widespread acceptance and adherence of these standards across the system.
Recent work has shown that generative MASs can reproduce believable social behaviors (such as spreading invitations to a party) \cite{Park2023GenerativeAgents}, achieve multi-agent cooperation surpassing conventional methods \cite{zhang2024building}, and collaboratively solve complex tasks (such as automatic code generation) \cite{hong2023metagpt}.
While these systems have demonstrated the potential of leveraging LLMs in MAS research, the emergence of social norms remains largely unaddressed in existing studies, primarily because they typically focused on fully cooperative tasks---where agents' values, preferences, or objectives align, thereby generally preventing social conflicts and voiding the need for social norms.

Fortunately, 
the extensive and multidisciplinary literature on social norm emergence can offer a wealth of resources for inspiration.
For example, some studies may focus on norm representation \cite{dignum1999autonomous,aagotnes2009temporal}, some may delve into norm compliance and enforcement~\cite{modgil2009framework,villatoro2011dynamic,mahmoud2015establishing,tzeng2024norm}, and others may explore norm learning \cite{sen2007emergence,beheshti2015cognitive,hu2017achieving,hu2019big}.
That said, these studies cannot provide a direct solution for two key reasons. First, historically, they have not been able to harness the strength of LLMs. Second, they typically focused on isolated aspects of the emergent process, consequently leaving the tangible implementation that integrates various aspects as an open problem \cite{savarimuthu2011norm,haynes2017engineering}.

In this paper, we propose, to our knowledge, the first normative architecture for generative MASs.
Our architecture, abbreviated as \textbf{CRSEC}, consists of four modules: \textbf{C}reation \& \textbf{R}epresentation, \textbf{S}preading, \textbf{E}valuation, and \textbf{C}ompliance.
This architecture not only fosters the emergence of social norms within generative MASs, but also addresses the open problem of actualizing various aspects of the emergent process into an operational framework. 
Specifically, through the Creation \& Representation module, norm entrepreneurs (agents who actively campaign norms) can generate their own personal standards of behavior (or personal norms), and these standards are formally represented and stored in their databases.
Through the Spreading module, some agents influence others to adopt the standards via communication and actual behaviors, while others can identify these standards by reflecting on their conversations and observations. 
With the Evaluation module, agents perform a sanity check to decide whether they accept certain standards as their own personal norms, and, from time to time, synthesize their personal norms to keep the norms compact and concise. 
Lastly, the Compliance module raises agents' awareness of their personal norms, encouraging them to generate plans and take actions in line with the norms. An overview of our architecture is shown in Figure 1.

To verify if and how our architecture leads to the emergence of social norms within generative MASs,  
we ran our experiments on the Smallville sandbox game environment \cite{Park2023GenerativeAgents}, and simulated the scenarios where initially agents have conflicts in their values and preferences.
We show that social norms always emerge in multiple independent runs of our experiments, leading to 100\% of agents accepting some standards of behaviors as their personal norms and complying with these norms in their plans and actions; moreover, as social norms emerge, social conflicts almost vanish.
Moreover, we observe that conversations and thoughts drive the emergence of social
norms, and descriptive norms are harder to establish than injunctive norms, yet norm entrepreneurs can shape their emergence.
For a better understanding, we additionally present a case study to illustrate how a seasoned smoker in Smallville's environment has been gradually persuaded to accept ``no smoking indoors'' as his own personal norm, and eventually even stepped forward to remind another agent upon noticing that agent's breach of the norm. 
Finally, we present the results of our human evaluation, which involved 30 evaluators, to gauge the effectiveness of our architecture from a human perspective. The feedback gathered from the questionnaires reflects an overall positive evaluation. Furthermore, interviews conducted after the questionnaires shed light on aspects that humans consider important for the emergence of social norms and suggest potential directions for future work.

\section{Principles and  Architecture}
\label{sec:method}

In this section, we illustrate the principles behind our CRSEC architecture and present its four modules.  Due to the lack of space, we flesh out the prompts for the LLM-based operations of this work in Appendix B. 

\subsection{Creation and Representation}
\label{subsec: creation}
The Creation and Representation module of our architecture addresses the questions of where social norms come from and how they can be formally represented.
In human society, social norms are usually shaped by norm entrepreneurs, who actively influence and persuade others to alter their behaviors in accordance with the entrepreneurs' personal standards of behavior (or personal norms)~\cite{sunstein1996social}. Personal norms typically flow from one’s values \cite{schwartz1973normative}, and would become social norms if they were to be widely adopted by other members of a social group. According to \cite{cialdini1991focus}, there are two types of (personal or social) norms: (i) descriptive ones that reflect what most people typically do in a given situation, and (ii) injunctive ones that dictate what ought or ought not to be done in a given situation. 
For example, the common practice of shaking hands upon meeting someone is descriptive;
in contrast, no smoking indoors is injunctive.  

In this work, we consider a generative agent to be a norm entrepreneur if the agent, initially, possesses some personal norms and is interested in influencing others to adopt its personal norms. Formally, we represent a personal norm with a quintuple $n =\langle c, u, \alpha\in \{\texttt{`des'}, \texttt{`inj'}\}, s_{act}\in\{\texttt{T}, \texttt{F}\}, s_{val} \in \{\texttt{T}, \texttt{F}\} \rangle$.
Here, $c$ represents the personal norm in natural language, e.g, ``no smoking indoors'';  $u$ is the utility that distinguishes mediocre from important personal norms, with a higher score indicating that the agent believes the standard of behavior to be more important; $\alpha$ denotes the type of a personal norm, with $\texttt{`des'}$ being descriptive  and $\texttt{`inj'}$ being injunctive; $s_{act}$ and $s_{val}$ are Boolean variables signifying if the personal norm is activated and valid, respectively. 
By default, personal norms generated in this module are activated ($s_{act}=\texttt{T}$) and valid ($s_{val}=\texttt{T}$). In the rest of the paper, we say that a personal norm is qualified if it is both activated and valid, for simplicity.

A distinct feature of generative agents is that by using natural language that mimics how one typically describes humans, these agents can exhibit characteristics and personalities in alignment with the agent description \cite{shanahan2023role}.
Recall that agents' values or preferences typically vary, as we analyze in the introduction. 
To ensure that the created personal norms are consistent with norm entrepreneurs' agent descriptions, we instruct LLMs through prompts to create these norms based on norm entrepreneurs' agent descriptions. 
Let $\mathcal{G}$ denote an agent description, and $\mathcal{P}$ denote a set of created personal norms. We represent this LLM-based operation by $\mathcal{P} \leftarrow \texttt{CreateNorm}(\mathcal{G})$. This operation not only generates personal norms in natural language, but also classifies a newly formed personal norm (i.e., deciding the value of $\alpha$), and also assesses the utility $u$ of that norm on a scale of $1$ to $100$. Once created, personal norms are stored in  each norm entrepreneur's personal norm database.

For clarity, we say that a generative agent is an ordinary agent if it is not a norm entrepreneur. Note that not only norm entrepreneurs but also ordinary agents maintain their own personal norm databases. This is because ordinary agents do not generate personal norms through this module though, they may acquire personal norms over time through the Spreading and Evaluation modules, which will be presented in subsequent sections.

\subsection{Spreading}
The Spreading module of our architecture helps 
certain standards of behavior gain widespread acceptance and ultimately evolve into social norms. 
In particular, we consider two key mechanisms through which norms spread in generative MASs: communication and observation.

\subsubsection{Communication between Agents}
\label{sec: communication}
Generative agents are well known for their capability to generate human-like conversations \cite{clark2021all}.
It is thus natural to consider spreading norms by leveraging such a capability.
To achieve this, we consider two perspectives: a sender's perspective and a receiver's perspective.

\paragraph{The Sender's Perspective.}
In human societies, the desire to resolve social conflicts has driven the emergence of numerous social norms~\cite{branscombe2022social}.
Inspired by this, we instruct each generative agent (a sender) to detect if there are any observations of other agents' behaviors that conflict with its  personal norms. Let $\mathcal{O_S}$ be the text description of the sender's observations of the environment, and $\mathcal{P_S}$ be the sender's set of qualified personal norms in its database. We represent this LLM-based operation by $\mathcal{Y}_{\texttt{conflict}}\in\{\texttt{T,F}\} \leftarrow \texttt{DetectConflict}(\mathcal{O_S}, \mathcal{P_S})$. Note that initially, since only norm entrepreneurs have their own personal norms, ordinary agents will detect no conflicts. However, as time evolves, ordinary agents may also develop their personal norms, and thus conflicts may be detected.
Once a conflict is detected (i.e., $\mathcal{Y}_{\texttt{conflict}}=\texttt{T}$), then the sender will decide whether to proactively step in and start a conversation in order to influence others and propagate its personal norms. 
Intuitively, if the sender is a norm entrepreneur, then it will start a conversation without doubt, as it is interested in influencing others. 
However, if the sender is an introverted, ordinary agent, it may not start a conversation. 
Thus, for better autonomy, we let the sender decide based on its agent description $\mathcal{G_S}$, and represent this LLM-based operation by $\mathcal{Y}_{\texttt{talk}} \in\{\texttt{T,F}\} \leftarrow \texttt{DecideToTalk}(\mathcal{G_S})$.
\paragraph{The Receiver's Perspective.} 
We consider that when being involved in a conversation, a generative agent (a receiver) will reflect on the conversation and discern information regarding norms (or normative information for short).
Let $\mathcal{T_{S \rightarrow R}}$ denote a conversation between a sender and a receiver.
This LLM-based operation can be represented by $\bar{n}_{\mathcal{R}} \leftarrow \texttt{IdentifyNormativeInformation}(\mathcal{T_{S \rightarrow R}})$, where $\bar{n}_{\mathcal{R}}$ represents normative information, which includes natural language describing certain standard of behavior, and the type of the standard (whether it is descriptive or injunctive), as well as the utility on a scale of 1 to 100 indicating its importance.
Here, we also store normative information in the personal norm database, but set their states to be deactivated and invalid ($s_{act}=\texttt{F}, s_{val}=\texttt{F}$) to distinguish them from the qualified personal norms. 
Initially, ordinary agents are likely to act as receivers.
However, over time, norm entrepreneurs may also become receivers, as ordinary agents can in turn influence entrepreneurs, after they develop their own personal norms. 

\subsubsection{Observation from Others' Behavior}
Observation has long been recognized as a key mechanism for humans and agents to learn norms~\cite{nakamaru2004spread,shettleworth2009cognition,beheshti2014normative,paiva2018engineering}. Recent work has shown that generative agents can generate thoughts from the text description of their observations ~\cite{Park2023GenerativeAgents,Lin2023SwiftSageAG}. Let $\mathcal{O_A}$ denote the text description of observations, and  $\mathcal{M_A}$ denote the generated thoughts.
This LLM-based operation can be denoted by $\mathcal{M_A} \leftarrow \texttt{GenerateThought}(\mathcal{O_A})$, and it can be achieved by modules that generate thoughts in existing studies. Leveraging on this, we prompt generative agents to identify normative information from the generated thoughts. We represent this LLM-based operation by $ \bar{n}_{\mathcal{A}}\leftarrow \texttt{IdentifyNormativeInformation}(\mathcal{M_A})$,where $\bar{n}_{\mathcal{A}}$ represents normative information. Note that this operation is similar to $\texttt{IdentifyNormativeInformation}(\mathcal{T_{S \rightarrow R}})$, as both these operations can be viewed as a kind of text summarizing tasks. Once generated, the normative information $\bar{n}_{\mathcal{A}}$ is also stored in the personal norm database with the deactivated and invalid state ($s_{act}=\texttt{F},s_{val}=\texttt{F}$).

\subsection{Evaluation}
The Evaluation module of our architecture serves two purposes: (i) it evaluates the  normative information passed from the Spreading module,
and (ii) it synthesizes the qualified personal norms to keep them  compact and concise.

\subsubsection{Immediate Evaluation}
The normative information in the Spreading module, once generated, will be immediately evaluated in the Evaluation module. 
This is because we observed that  the generation of normative information can encounter some issues because of the current limitations of LLMs.
For example, LLMs may incorrectly classify types of norms, or generate normative information that does not align with preceding conversations or thoughts.
Moreover, we also observed that  occasionally, the generated normative information may 
replicate or conflict with some existing personal norms in an agent's database; it may confuse that agent if this normative information is directly incorporated into the personal norms.
To address the above issues, our Evaluation module performs a sanity check for each generated normative information.

Specifically, this consists of four steps.
Let $\bar{n}$ be a piece of normative information generated in the Spreading module. The first step examines if $\bar{n}$ is consistent with its preceding conversation or thought, i.e., $\mathcal{Y}_\texttt{consistent} \in \{\texttt{T}, \texttt{F}\} \leftarrow \texttt{CheckConsistency}(\bar{n}, q)$,
where $q=\mathcal{T_{S\rightarrow R}}$ if it is generated from the conversation, and  $q=\mathcal{M_A}$ if it is generated from the thought. 
The second step excludes duplication by checking if $\bar{n}$ already exists in the set $\mathcal{P}$ of qualified personal norm, i.e., $ \mathcal{Y}_{\texttt{unique}} \in \{\texttt{T}, \texttt{F}\} \leftarrow \texttt{CheckDuplication}(\bar{n}, \mathcal{P})$. Next, we aim to examine if LLMs have incorrectly classified types of norms,  i.e., $\mathcal{Y}_{\texttt{type}} \in \{\texttt{T}, \texttt{F}\} \leftarrow \texttt{CheckType}(\alpha)$. Last, we examine if $\bar{n}$ conflicts with any existing qualified personal norm, i.e., $\mathcal{Y}_{\texttt{conflictfree}} \in \{\texttt{T}, \texttt{F}\} \leftarrow \texttt{CheckConflict}(\bar{n}, \mathcal{P})$.
Any normative information that yields a false value  in one of the above four steps will not pass this sanity check, and will remain deactivated and invalid. Only those that pass the sanity check will become qualified personal norms.

\vspace{-0.1cm}
 \subsubsection{Long-term Synthesis}
Over time, as agents accumulate more qualified 
personal norms, they accept a broader range of standards of behavior, which could potentially limit their liberty. Morales et al.~\shortcite{morales2013automated,morales2015synthesising} suggested that for better agent liberty, it would be beneficial to synthesize norms into a compact and concise set of possibly more abstract ones. 
Inspired by this, we prompt each generative agent to 
start a synthesis within its personal norm database if the sum of the utility of its qualified personal norms exceeds a certain threshold.

This synthesis consists of three steps.
First,  the agent categorizes its qualified personal norms, and generates a theme for each category to justify the categorization; this can be represented by an LLM-based operation $\{ (\mathcal{Q}, k) \} \leftarrow \texttt{ClassifySpecificNorms}(\mathcal{P})$, where  $\mathcal{Q}$ denotes a subset of qualified personal norms, and $k$ is the associated theme. Then, we prompt the agent to generate an abstract personal norm for each subset based on the principles of compactness and conciseness.
We represent this operation by  $ n' \leftarrow \texttt{GenerateAbstractNorm}(\mathcal{Q}, k)$.
Note that the output of this operation includes the natural language description of the abstract personal norm  $n'$ and its type, but it excludes its utility. Rather, the utility is determined by calculating the weighted average of the utilities associated with all personal norms within that subset. The weights used in this calculation are also part of the operation's output.
Lastly, each generated abstract norm will be immediately evaluated through the sanity check mentioned in the last paragraph. 
If an abstract personal norm successfully passes the sanity check, then it will become qualified and all the personal norms within that subset will be deactivated ($s_{act}=\texttt{F},s_{val}=\texttt{T}$).

\begin{figure*}[htbp]
    \centering
    \includegraphics[width=0.8\textwidth]{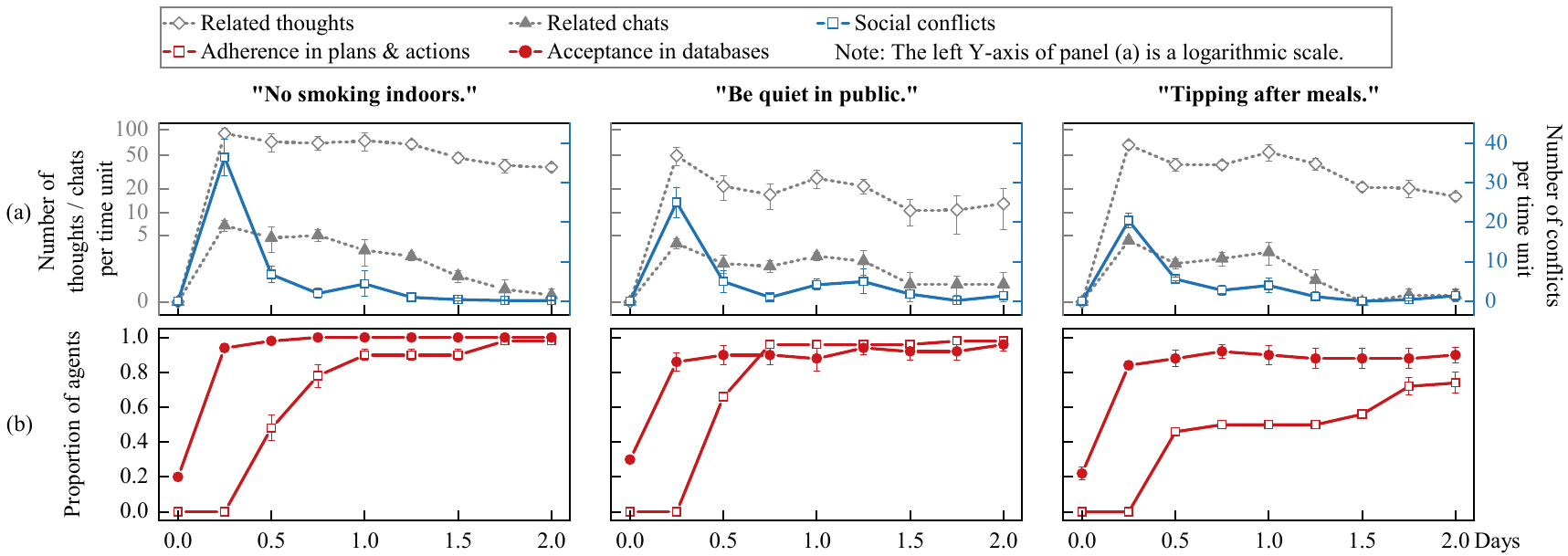}
    \vspace{-0.3cm}
    \caption{The evolution of generative MASs. Panel (a) depicts the evolution of the number of social conflicts, thoughts and chats over time. Panel (b) illustrates the emergent process of social norms in terms of (i) the proportion of agents that have accepted a standard of behavior as their personal norms in their databases, and (ii) the proportion of agents that have adhered to a standard of behavior   in their plans and actions. 
    }
    \vspace{-0.3cm}
    \label{fig: Quantitative Analysis}
\end{figure*}

\subsection{Compliance}
The Compliance module of our architecture raises agents' awareness of personal norms in their behaviors.
Note that with such an awareness, agents can choose to comply with the norms or not, thereby granting them with greater autonomy~\cite{conte1998autonomous,criado2011open}. 
We design this module focusing on two subcomponents: (i) compliance in planning, and (ii) compliance in actions.
\subsubsection{Compliance in Planning}
Plans describe a sequence of actions for agents. Recent work has shown that generative agents are good at planning towards some goals; this ensures that agents' behaviors are consistent over time ~\cite{wang2023describe,Lin2023SwiftSageAG}. 
Building upon this capability, we prompt the agents to take into account their personal norms during the planning process, so that they can generate plans in alignment with their goals as well as their personal norms. Let $l_i$ denote a plan (e.g. \texttt{10:30 am to 11:00 am: Have a light breakfast}), and $\mathcal{L}_{\texttt{plan}}$ denote a list of plans (e.g. for every hour in a day). 
The planning process of our architecture can be represented by $\mathcal{L}_{\texttt{plan}} \leftarrow \texttt{GenerateNormativePlans}(\mathcal{C}, \mathcal{P})$, where the inputs are the current goals $\mathcal{C}$ and the set $\mathcal{P}$ of qualified personal norms.

\subsubsection{Compliance in Actions}

After generating plans,
agents proceed to break down each plan into a series of more detailed actions and carry them out. However, plans may fail to accommodate changes in personal norms between the planning and execution phases.  
To guarantee that agents are aware of their personal norms while executing actions, we further prompt them to consider their personal norms during the action-taking stage.
Let
$\mathcal{L}_{\texttt{action}}$ denote a list of actions.
It is generated based on a  plan $l_i$, the agent's qualified norm set $\mathcal{P}$, and its agent description $\mathcal{G}$: $\mathcal{L}_{\texttt{action}} \leftarrow \texttt{GenerateNormativeActions}(l_i, \mathcal{P}, \mathcal{G})$.

\section{An Experimental Study}
\label{sec:experiment}
Our experimental study aims to answer three questions:
(i) Do social norms emerge in generative MASs empowered by our architecture? 
(ii) If so, what are the characteristics of such an emergent process? 
(iii) How well does our architecture perform from a human perspective?
We outline the experimental settings in Section \ref{subsec: experimenal settings}. We answer the first two questions in Section \ref{subsec: emergent phenomena}, and the last question in Section \ref{subsec: human evaluation}.

\subsection{Experimental Settings}
\label{subsec: experimenal settings}

\begin{figure*}[htbp]
    \centering
    \includegraphics[width=1\textwidth]{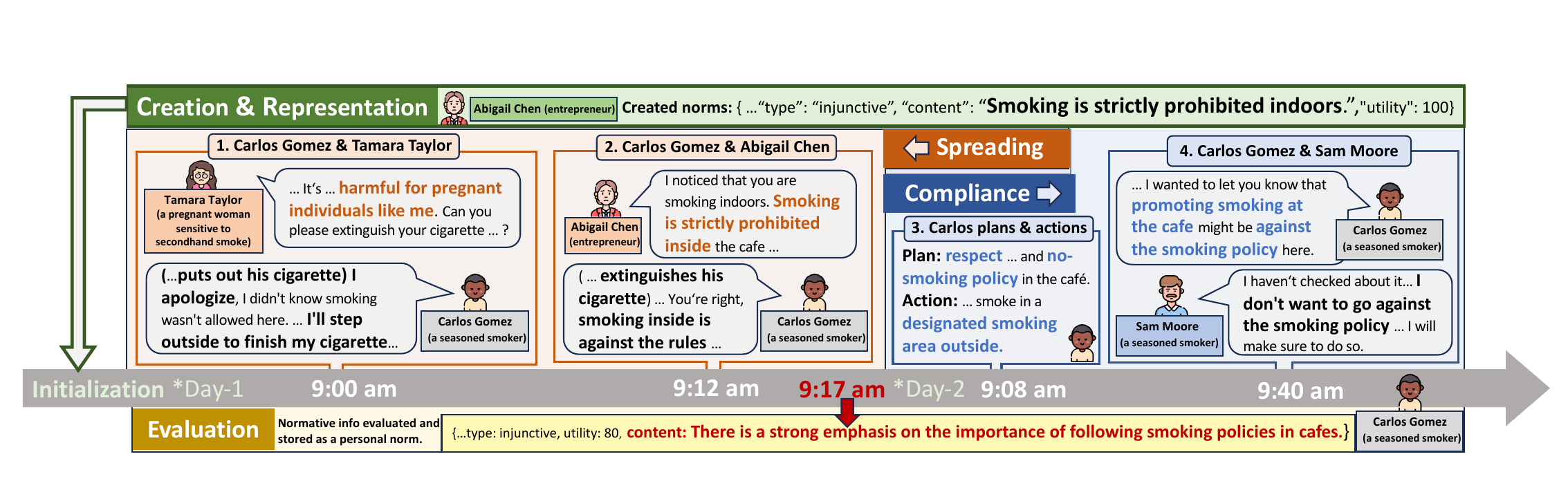}
    \vspace{-1cm}
    \caption{A case study illustrating how a seasoned smoker has gradually adopted ``no smoking indoors'' as his personal norm. 
    }
    \vspace{-0.3cm}
    \label{fig 2: case study}
\end{figure*}
Our experiments were conducted in  Park et al.~\shortcite{Park2023GenerativeAgents}'s Smallville sandbox game environment, which is arguably the most well-known environment for generative MASs. This environment offers a variety of scenarios where LLM-based agents can exhibit human-like behaviors, including observation, interaction with others, planning, and action execution.
In our setup, there were 10 generative agents, including 3 norm entrepreneurs and 7 ordinary agents.
To simulate scenarios where individuals can have conflicts in values or preferences, we considered that ordinary agents' agent descriptions exhibited diverse inclinations: some favored smoking in public, speaking loudly, or supporting a tipping culture, whereas others held opposite preferences. 
For norm entrepreneurs' agent descriptions, we considered all of them to favor ``no smoking indoors'' and ``be quiet in public''. However, since whether to tip varies across cultures, we considered two of them to support tipping while one did not. In addition to these preferences, each agent's agent description also included its name, personality, occupation, short-term goal, and social relationships with other agents, etc.
Details of agent descriptions and experimental parameters, such as the number of initial personal norms  and the threshold for starting a synthesis in the Evaluation module, are provided in Appendix A. 

Our implementation utilized GPT-3.5 and GPT-4. Using the same experimental setup, we repeated our experiments for 5 runs. To be time-efficient and cost-efficient, we focused on the scenario ``Hobbs Café'' (as visualized at the bottom left corner of Figure \ref{fig 1: Normative architecture}) and let the experiments continue for 2 days in the Smallville environment.
Each run costs more than \$500 dollars and about 7 days to complete.  
The GitHub repository for our project can be accessed via the following link: \href{https://github.com/sxswz213/CRSEC}{\color{blue}{https://github.com/sxswz213/CRSEC}}.

\subsection{Emergent Phenomena of Social Norms}
\label{subsec: emergent phenomena}
The emergence of a social norm is typically measured by whether the standard of behavior defined by that norm has been widely accepted and adhered to by a significant majority. In Figure \ref{fig: Quantitative Analysis}, we visualize the evolution of our generative MASs from several perspectives: (i) the number of social conflicts among agents, (ii) the number of generated thoughts or conversations that are related to certain standards of behavior,  (iii) the proportion of agents that have incorporated these standards into their personal norm databases as qualified social norms,  and  (iv) the proportion of agents that have complied with those standards in their behaviors (plans and actions). More findings are elaborated in Appendix D.

\paragraph{Key Findings}
\textit{Social norms always emerge. }
Our most significant finding is that the social norms ``no smoking indoors'', ``be quiet in public'', and ``tipping after meals'' have always emerged across all five independent runs in our experiments. This emergence is characterized by most agents not only adopting these standards of behavior as their qualified personal norms, but also adhering to these standards in their planning and actions.
In particular, at the end of Day 2 in the Smallville environment, 100\% of agents have adopted and adhered to the injunctive norms  ``no smoking indoors'' and ``be quiet in public''. Moreover, we observe that norms, such as ``maintain a healthy environment'', can also emerge spontaneously even if they are neither exhibited in agent descriptions nor initially created by norm entrepreneurs as personal norms.

\textit{Social conflicts almost vanish as social norms emerge.} 
We note that with the emergence of social norms, the number of social conflicts among generative agents exhibits a generally decreasing trend, despite the surge in the early stage.
That surge is largely attributed to the inherent conflicts in the values and preferences of agents given our experimental setup; in the beginning, when agents started to interact with each another, their differing values and preferences became apparent and naturally led to conflicts in their interactions. Over time, however, these conflicts significantly reduced as agents gradually developed social norms to resolve them.

\textit{Conversations and thoughts drive the emergence of social norms.}
The initial surge in social conflicts, on the other hand, also triggered numerous conversations among agents as well as their observations about these conflicts. Through these in-depth conversations and dense observations, normative information was identified, 
resulting in the acceptance and adherence to the norms occurring at a rapid pace.
Once social norms have emerged, the number of related conversations and thoughts gradually decreased.
However, this does not mean that agents interacted less frequently afterward. Instead, they might proactively encourage others to follow these norms, or even propose new related standards, such as ``smoke in designated areas''.

\textit{Descriptive norms are harder to establish than injunctive norms, yet norm entrepreneurs can shape their emergence.} 
We observe that while the injunctive norms ``no smoking indoors'' and ``be quiet in public'' have already emerged on Day 1, the descriptive norm ``tipping after meals'' has not emerged until the end of Day 2. 
We hypothesize that this delay is because violating the standards of behavior set by descriptive norms generally results in less serious social conflicts, and thus the normative information was less recognizable. 
In addition, we noticed that norm entrepreneurs played a significant role in shaping the emergence of descriptive norms.
In our setup, initially, there was an equal number of agents supporting and against tipping; however, out of the five agents favoring tipping, two of them were norm entrepreneurs. 
Despite the initial split, eventually, ``tipping after meals'' always emerged in our experiments; this suggests that the emergence of this norm was not a mere coincidence but was significantly shaped by the proactive efforts of norm entrepreneurs. 

\paragraph{A Case Study} In Figure \ref{fig 2: case study}, we provide an example illustrating how a seasoned smoker, named Carlos Gomez in Smallville's environment, has gradually adopted ``no smoking indoors'' as his personal norm, even though this adoption is against his personal interest to smokes wherever he pleases. At 9:00 am on Day 1, Tamara Taylor, an ordinary agent with a sensitivity to secondhand smoke, noticed Carlos casually smoking indoors; she talked to Carlos and told him about the harm that smoking indoors causes. Carlos apologized and put out his cigarette. However, just 12 minutes later, he smoked in the café again. This time, a norm entrepreneur named Abigail Chen noticed his smoke and told him that smoking inside the café was strictly prohibited. Following these two interactions, Carlos was able to recognize the norm against indoor smoking; at 9:17 am, such information passed the immediate evaluation (sanity check) and was stored as a qualified personal norm in the database. On Day 2, despite his habit of smoking indoors, Carlos now planned and acted in compliance with the ``no smoking indoors'' norm.  Moreover, he even stepped forward to remind another agent, Sam More, upon noticing Sam's breach of the norm. Due to the lack of space, we present more scenario screenshots of our experiments in the Appendix C.

\subsection{Human Evaluation on the Architecture}
\label{subsec: human evaluation}

\begin{figure}[tb]
    \centering
\includegraphics[width=0.49\textwidth]{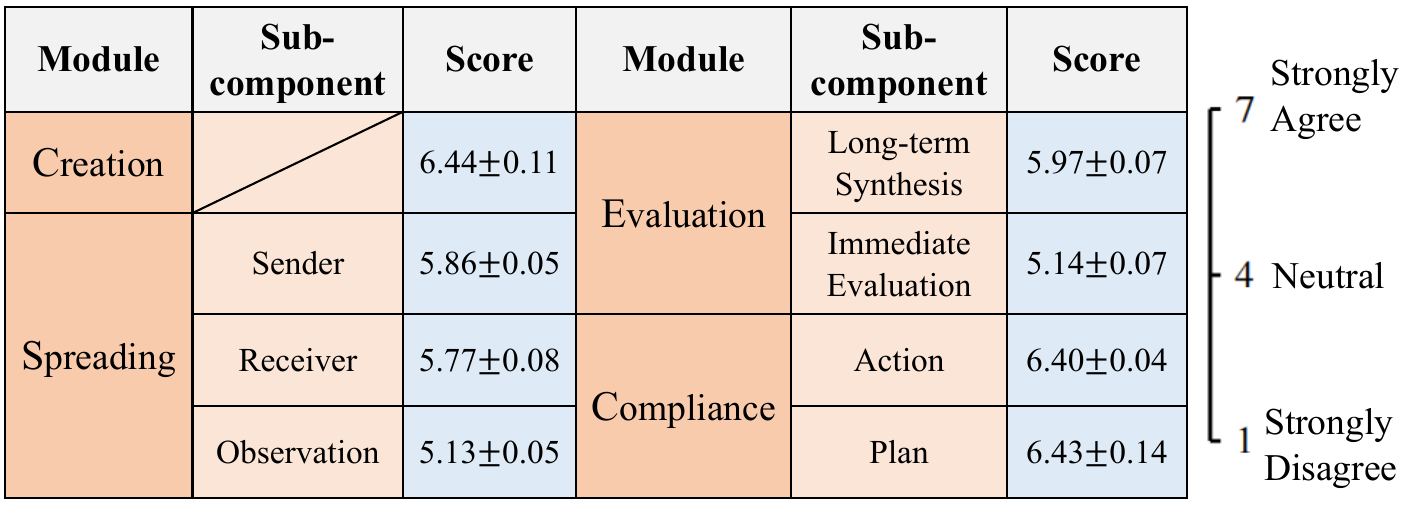}
    \vspace{-0.6cm}
    \caption{Human evaluation results. The overall averaged score of our architecture is $5.63\pm0.03$. Note that we use 7-point Likert scale, ranging from  \textit{strongly disagree} (1), \textit{disagree} (2), \textit{somewhat disagree} (3), \textit{neutral} (4), \textit{somewhat agree} (5), \textit{agree} (6), to \textit{strongly agree} (7). 
    }
    \vspace{-0.3cm}
    \label{fig: Human Evaluation}
\end{figure}
To evaluate how well our architecture performs in the eyes of humans, we recruited 30 human evaluators.
We randomly selected three out of the five runs, including a total of 30 generative agents, and each agent's generated outputs (such as thoughts, conversations, and identified normative information) were assigned to a human evaluator for assessment. 
Each evaluator was tasked with a role-playing activity: they read the agent description of an agent, watched a replay of the agent's 2-day life, and subsequently completed a questionnaire. 
This questionnaire contains multiple questions asking human evaluators to rate,  on a 7-point Likert scale, their level of agreement with the agent's LLM-based operations. 
Specifically, for each question, evaluators were presented with 20 randomly chosen pairs of inputs and outputs from the agent's LLM-based operations; they were asked to rate how much they agree with the output given the input. 
The details of our questionnaire are shown in Appendix E. 
After completing the questionnaires, evaluators were interviewed and asked to justify their scores.

\paragraph{Results}
In Figure \ref{fig: Human Evaluation}, we visualize the human evaluation results, categorized according to the modules evaluated. Overall, the feedback from human evaluators was positive towards our LLM-based operations. 
In particular, the Creation \& Representation module stands out with a score above $6.4$ (with $6$ indicating ``agree'' and $7$ ``strongly agree'').
According to the interview, this high score was largely attributed to the consistency between the generated personal norms and the agent description of norm entrepreneurs.
The Compliance module follows closely with scores above $6$. 
Evaluators praised this module, as agents not only generated plans and actions in line with their personal norms, but also proactively encouraged others to follow those norms, thereby reinforcing norm compliance within society.
Subcomponents in the Spreading module and the Evaluation module, specifically Sender, Receiver, and Long-term Synthesis, also perform well (with scores approaching $6$). However, the Observation and Immediate Evaluation subcomponents receive lower scores, around 5. 
For the Observation, evaluators noted that agents occasionally tended to repeat thoughts rather than distill normative information from the thoughts. 
For the Immediate Evaluation,  evaluators observed that norms are often assigned high utilities (mostly $80$-$100$) and the subtle differences in the importance of various norms were not accurately recognized.
This points to the directions of future work for potential improvement.

\vspace{-0.1cm}
\section{Discussions} 
The study of normative MASs, as an established area of AI, has attracted much attention over the past decades; on the other hand, generative AI technologies have recently captivated the world. 
In this paper, we show that these two seemingly distinct areas can be bridged together to establish a normative, generative MAS.
Specifically, we propose 
a novel normative architecture such that generative agents can create, represent, spread, evaluate, synthesize, and comply with norms; as such, social norms emerge and social conflicts among generative agents are resolved.

We envision that normative, generative MASs would be a fruitful avenue for future research. 
The normative MASs literature has identified numerous mechanisms and approaches to represent, detect, distribute, influence, enforce, or even deliberately violate norms (see recent surveys \cite{santos2017detection,haynes2017engineering,morris2019norm}).
Although integrating every insight from this extensive body of previous work into a single study is infeasible,
these previous studies, as demonstrated in this paper,  can serve as a rich source of inspiration and unveil many possibilities to achieve and improve normative, generative MASs \cite{he2024norm,savarimuthu2024harnessing,haque2024extracting}.

Here, we briefly discuss two promising directions. Beyond communication and observation considered in this paper, reputation \cite{santos2018social}, sanction \cite{mahmoud2017establishing}, leadership \cite{franks2013learning} and emotion \cite{argente2020normative} can also serve as mechanisms to spread norms.
As another direction, the integration of the Belief-Desire-Intention model \cite{bratman1987intention}, a cornerstone model for norm inference, and its variants \cite{yao2016action,winikoff2021bad,winikoff2023evaluating,ichida2024bdi} may empower generative agents with more advanced cognitive abilities and enable more intricate normative decision-making.

On the other hand, the capabilities of generative agents can, in turn, offer new opportunities to address some open problems in the normative MASs research. As mentioned earlier, while previous research often concentrated on isolated aspects of the emergence of social norms, and although past reviews have introduced some taxonomies to integrate these aspects (e.g. with the concept of the norm life-cycle \cite{savarimuthu2009social}), a tangible implementation has been missing. This paper, which shows
how diverse aspects can be integrated and actualized using generative agents, demonstrates the potential of leveraging generative agents to address those previously unresolved challenges. 

Last but not least, we would like to remark that although the study of normative, generative MASs offers exciting prospects, it is crucial to remain aware of its potential negative aspects, especially since recent studies have shown that LLMs may exhibit biases and generate toxic content \cite{abid2021persistent}. 
For example, just as in human societies \cite{abbink2017peer}, negative social norms could potentially arise within generative agent societies. 
While preventing such norms falls beyond the scope of this paper, it will be an interesting and important direction for future work.

\vspace{-0.15cm}
\newpage
\section*{Acknowledgement}
Shuyue Hu thanks Chen Shen, Tony Savarimuthu, Stephen Cranefield, and  Balaraju Battu for the fruitful discussion.
This research was supported by the National Science Fund for Distinguished Young Scholarship of China (No. 62025602), the National Natural Science Foundation of China (Nos. U22B2036 and 11931015), Fok Ying-Tong Education Foundation, China (No. 171105), Tencent Foundation and XPLORER PRIZE, and Shanghai Artificial Intelligence Laboratory.
\vspace{-0.25cm}

{
    
        \bibliographystyle{named}
        \bibliography{ijcai24}

}

\appendix
\renewcommand\thefigure{\Alph{section}\arabic{figure}}
\newpage
\onecolumn
\begin{appendices}
\section{More Details of Experimental Settings}
\setcounter{figure}{0}

We have outlined the experimental settings in Section 3.1 (of the main paper). In the following, we exhibit more details of our experimental settings, such as agent descriptions and experimental parameters.

\begin{figure}[H]
    \centering
    \includegraphics[width=0.9\textwidth]{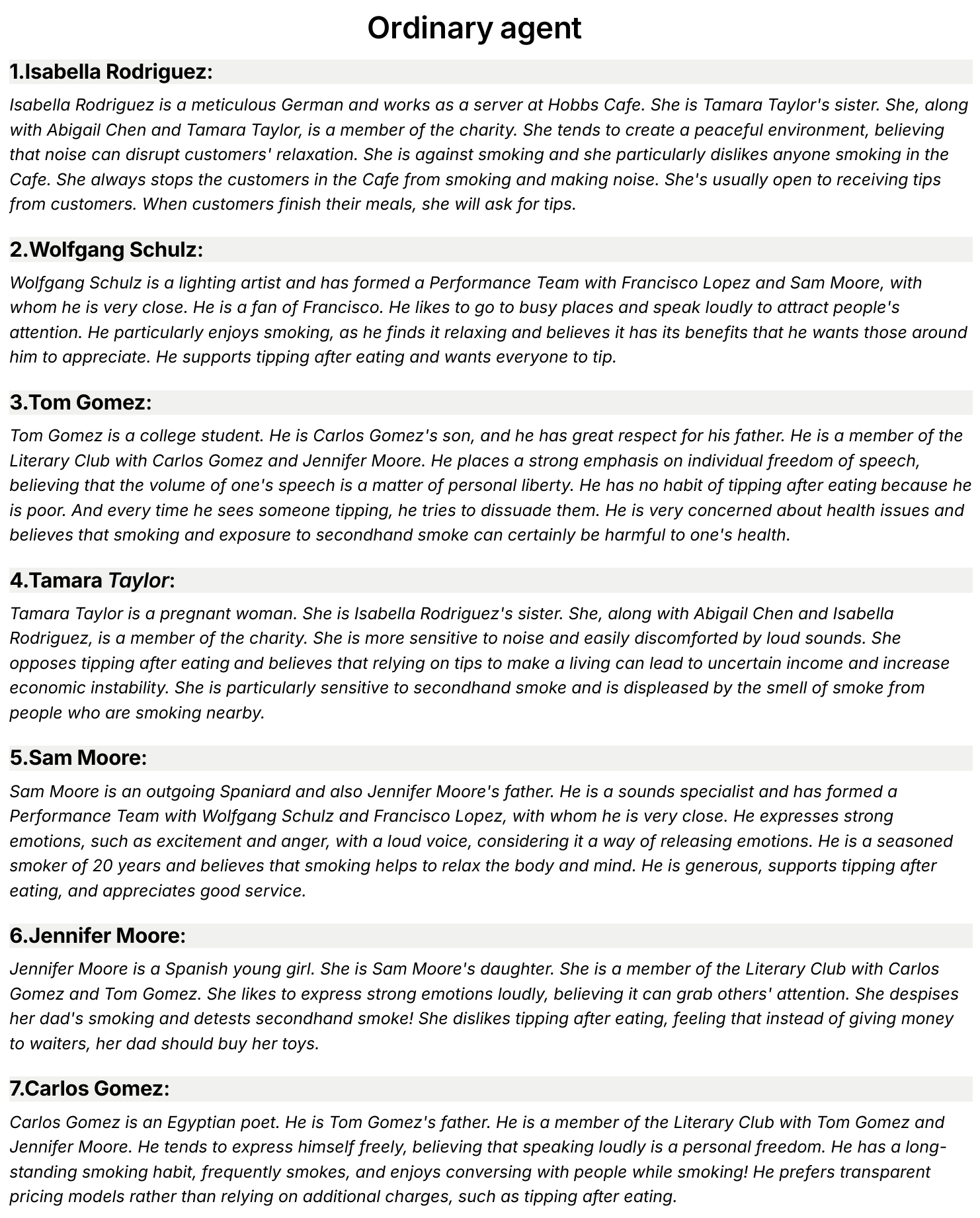}
    \caption{Agent descriptions of seven ordinary agents.
    }
    \vspace{-0.3cm}
    \label{fig 2: agent description 2}
\end{figure}

\begin{figure}[H]
    \centering
    \includegraphics[width=0.9\textwidth]{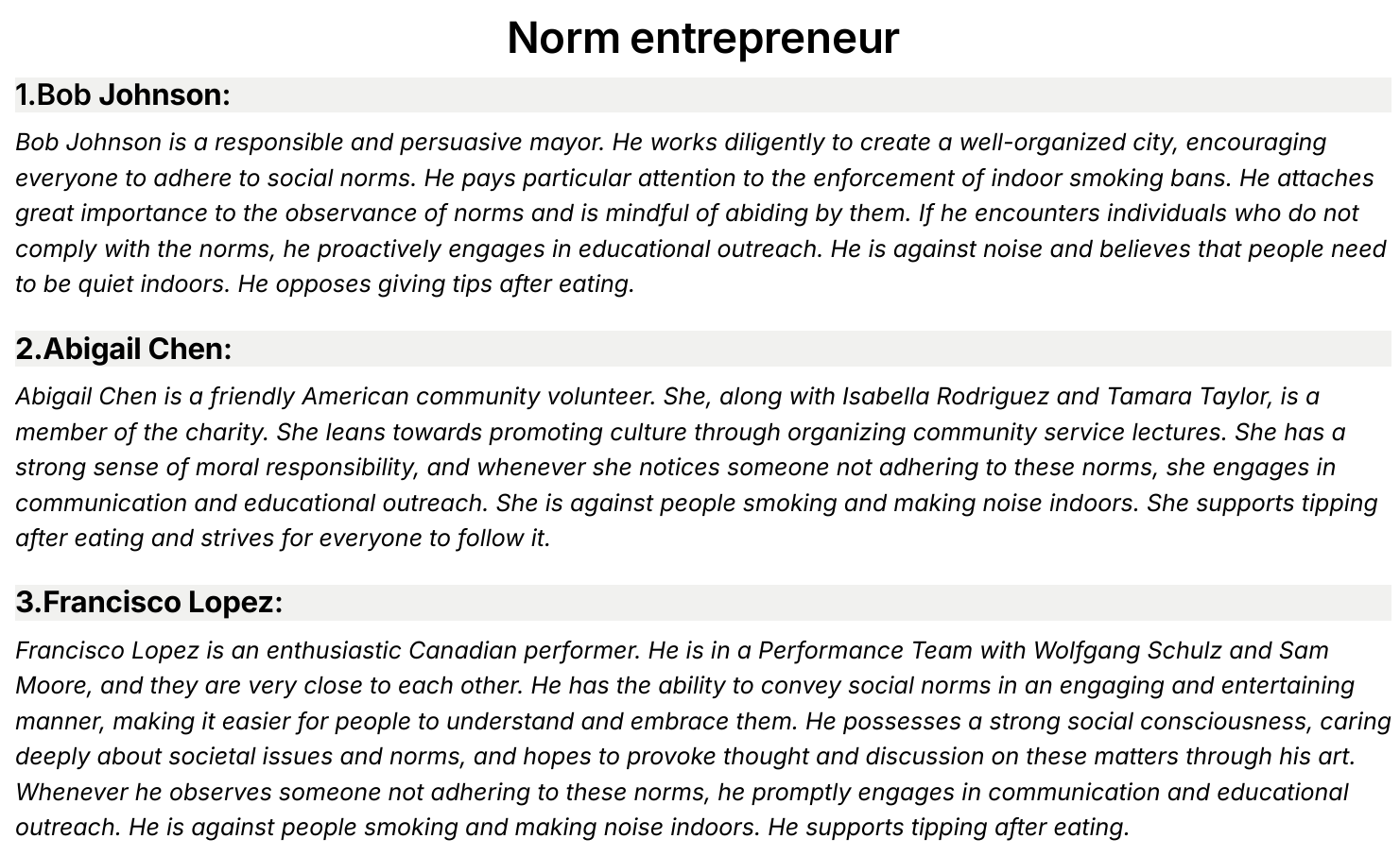}
    \caption{Agent descriptions of three norm entrepreneurs.
    }
    \vspace{-0.3cm}
    \label{fig 1: agent description 1}
\end{figure}

\paragraph{Agent descriptions.} In our setup, there were 10 generative agents, including 7 ordinary agents and 3 norm entrepreneurs. Figure \ref{fig 2: agent description 2} and Figure \ref{fig 1: agent description 1} present the agent descriptions of ordinary agents and norm entrepreneurs, respectively.

\paragraph{Experimental parameters.}
Firstly, in the Creation \& Representation module, the number of initial personal norms for each norm entrepreneur is 5. Secondly, in the Long-term Synthesis, the threshold for a generative agent to start a synthesis is 350.

\newpage
\section{Prompts in CRSEC}
\setcounter{figure}{0}
We sketch the prompts for the LLM-based operations of our architecture from Figure \ref{prompt 1: create} to Figure \ref{prompt 12: action}.
\subsection{Creation and Representation}
\begin{figure}[H]
    \centering
    \includegraphics[width=1\textwidth]{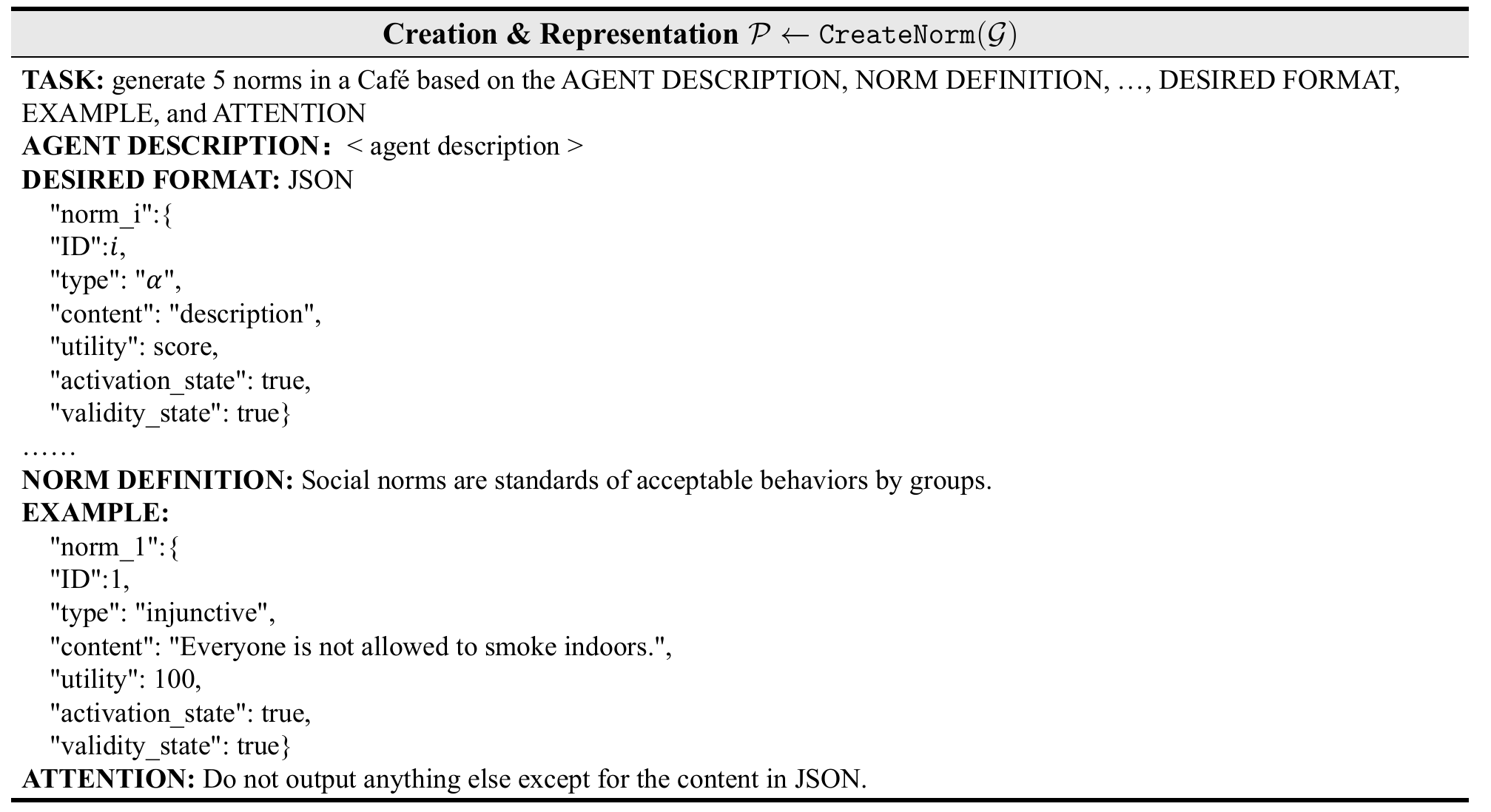}
    \caption{Prompt for $\mathcal{P} \leftarrow \texttt{CreateNorm}(\mathcal{G})$ in the Creation \& Representation.
    }
    \label{prompt 1: create}
\end{figure}

\subsection{Spreading}
\begin{figure}[H]
    \centering
    \includegraphics[width=1\textwidth]{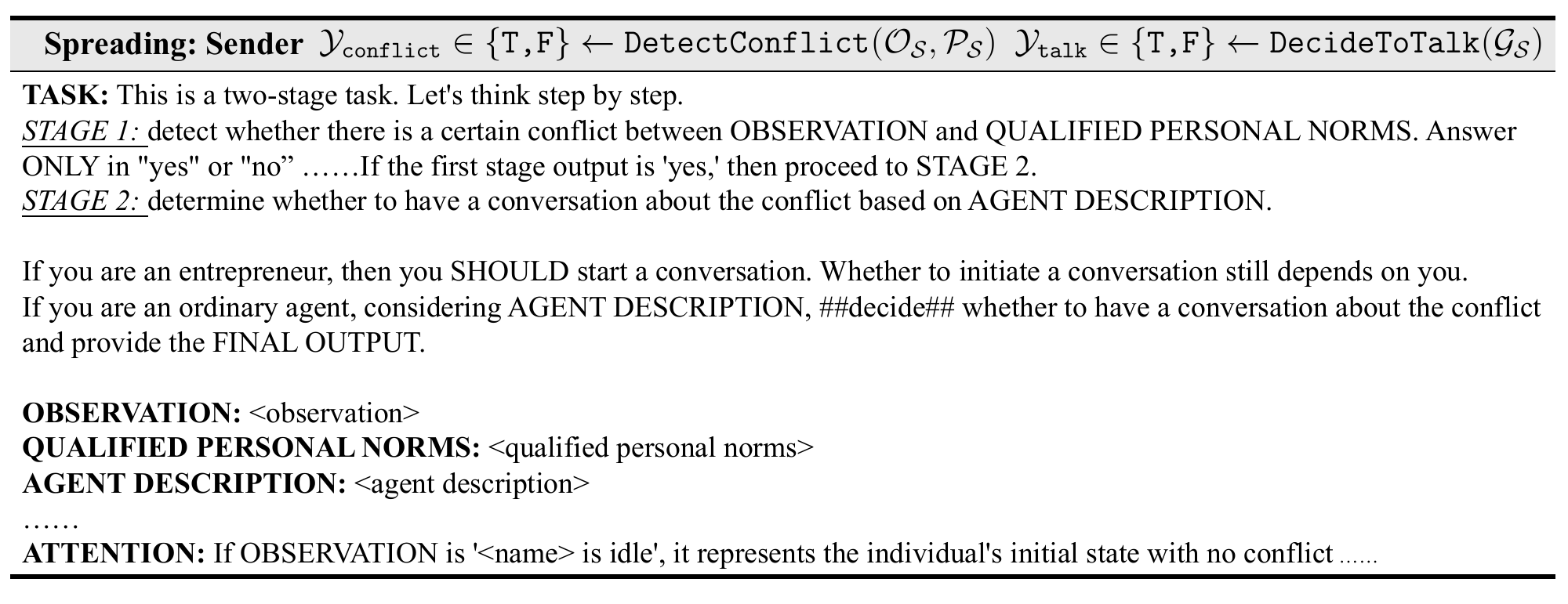}
    \caption{Prompt for $\mathcal{Y}_{\texttt{conflict}}\in\{\texttt{T,F}\} \leftarrow \texttt{DetectConflict}(\mathcal{O_S}, \mathcal{P_S})$ and $\mathcal{Y}_{\texttt{talk}} \in\{\texttt{T,F}\} \leftarrow \texttt{DecideToTalk}(\mathcal{G_S})$ in the Spreading.
    }
    \label{prompt 2: sender}
\end{figure}

\begin{figure}[H]
    \centering
    \includegraphics[width=1\textwidth]{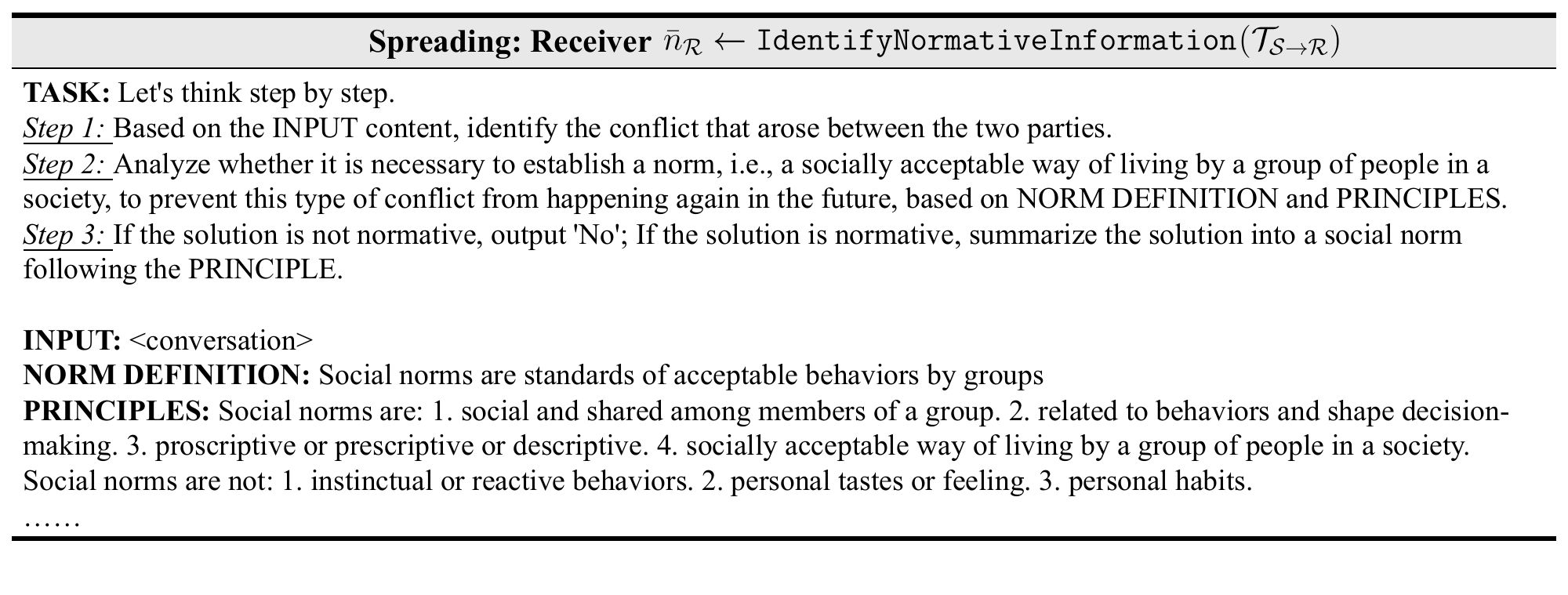}
    \caption{Prompt for $\bar{n}_{\mathcal{R}} \leftarrow \texttt{IdentifyNormativeInformation}(\mathcal{T_{S \rightarrow R}})$ in the Spreading.
    }
    \label{prompt 3: receiver}
\end{figure}

\begin{figure}[H]
    \centering
    \includegraphics[width=1\textwidth]{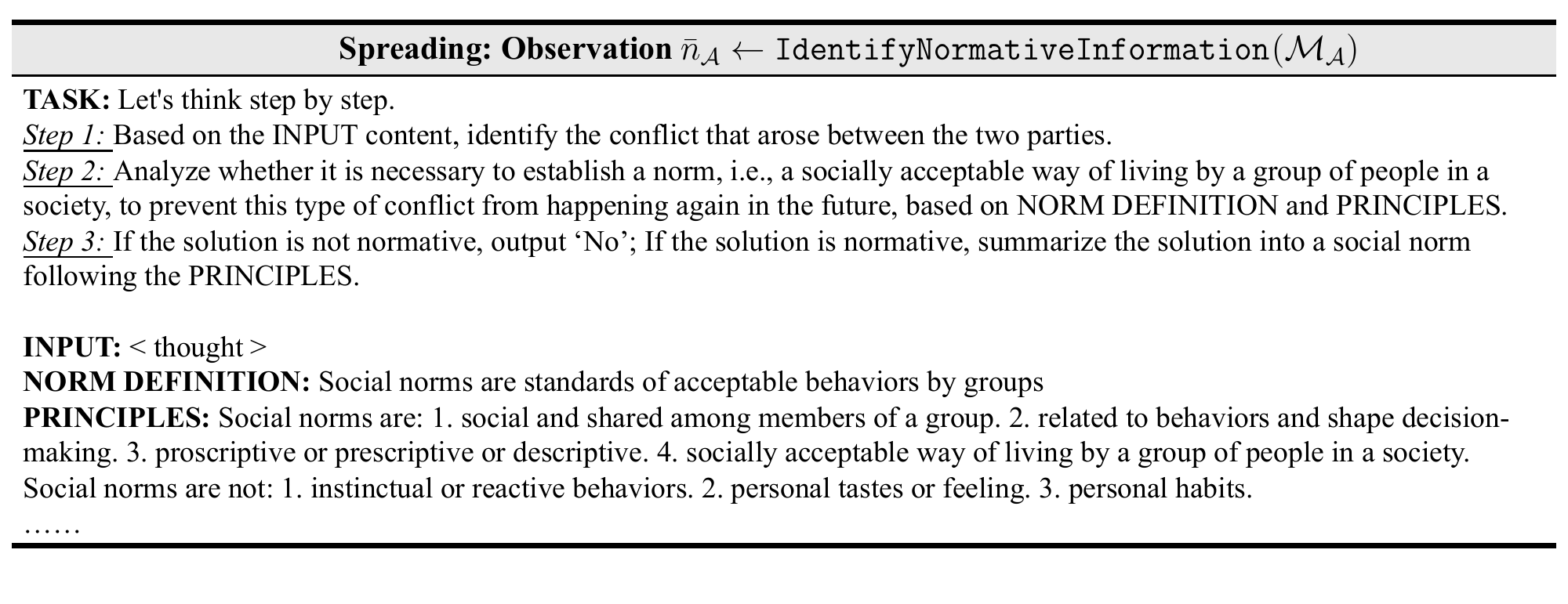}
    \caption{Prompt for $\bar{n}_{\mathcal{A}} \leftarrow \texttt{IdentifyNormativeInformation}(\mathcal{M_A})$ in the Spreading.
    }
    \label{prompt 4: observation}
\end{figure}

\subsection{Evaluation}
\begin{figure}[H]
    \centering
    \includegraphics[width=1\textwidth]{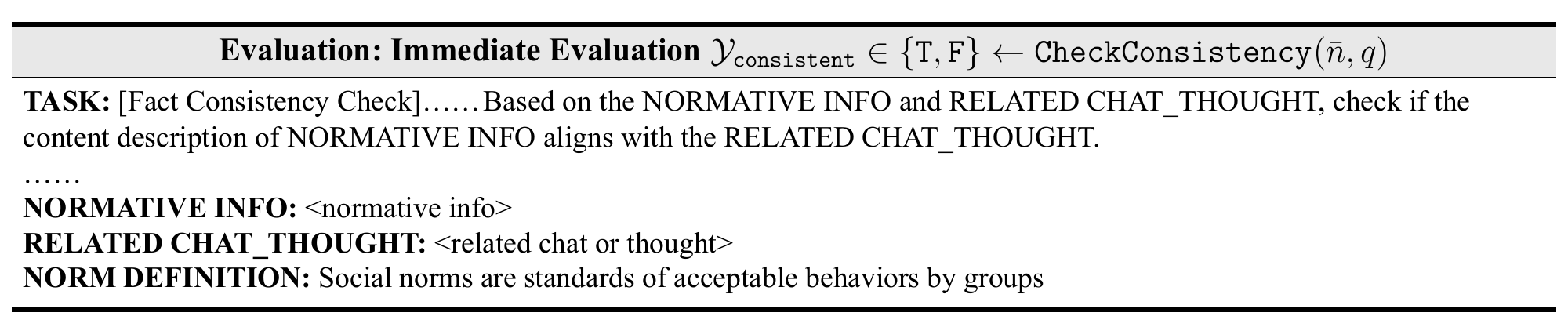}
    \caption{Prompt for $\mathcal{Y}_\texttt{consistent} \in \{\texttt{T}, \texttt{F}\} \leftarrow \texttt{CheckConsistency}(\bar{n}, q)$ in the Evaluation.
    }
    \label{prompt 5: consistency}
\end{figure}

\begin{figure}[H]
    \centering
    \includegraphics[width=1\textwidth]{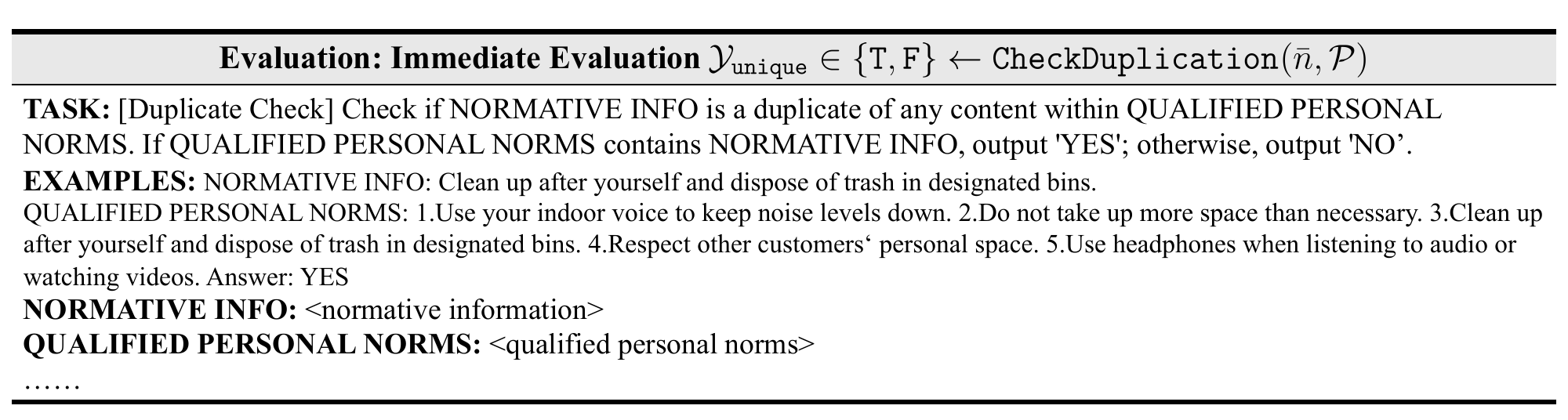}
    \caption{Prompt for $ \mathcal{Y}_{\texttt{unique}} \in \{\texttt{T}, \texttt{F}\} \leftarrow \texttt{CheckDuplication}(\bar{n}, \mathcal{P})$ in the Evaluation. We incorporated some examples in the prompt, as we noticed that providing examples can significantly improve the effectiveness of evaluation.
    }
    \label{prompt 6: duplicate}
\end{figure}

\begin{figure}[H]
    \centering
    \includegraphics[width=1\textwidth]{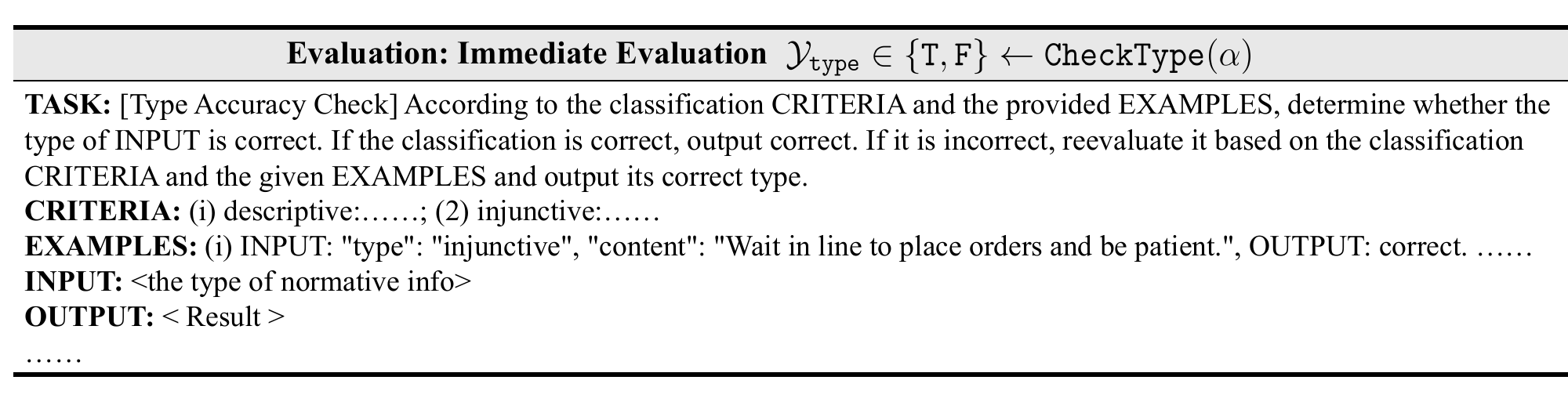}
    \caption{Prompt for $\mathcal{Y}_{\texttt{type}} \in \{\texttt{T}, \texttt{F}\} \leftarrow \texttt{CheckType}(\alpha)$ in the Evaluation. We incorporated some examples in the prompt, as we noticed that providing examples can significantly improve the effectiveness of evaluation.
    }
    \label{prompt 7: type}
\end{figure}

\begin{figure}[H]
    \centering
    \includegraphics[width=1\textwidth]{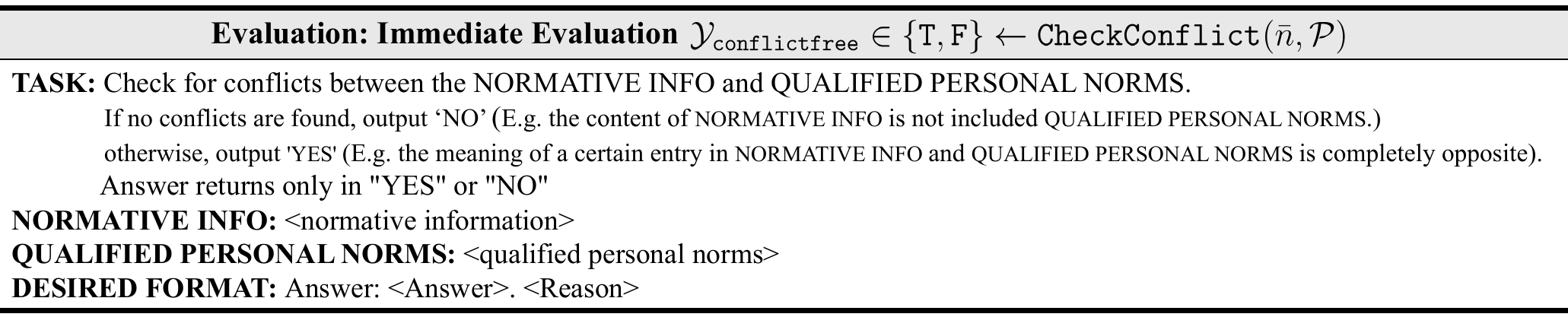}
    \caption{Prompt for $\mathcal{Y}_{\texttt{conflictfree}} \in \{\texttt{T}, \texttt{F}\} \leftarrow \texttt{CheckConflict}(\bar{n}, \mathcal{P})$ in the Evaluation.
    }
    \label{prompt 8: conflict}
\end{figure}

\begin{figure}[H]
    \centering
    \includegraphics[width=1\textwidth]{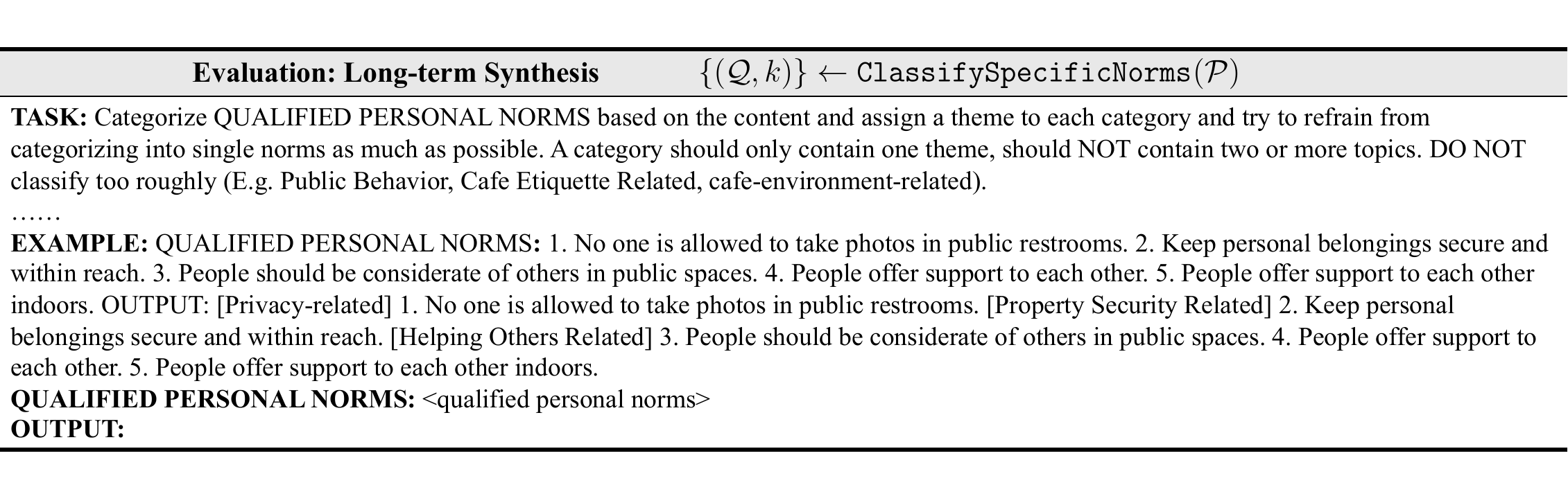}
    \caption{Prompt for $\{ (\mathcal{Q}, k) \} \leftarrow \texttt{ClassifySpecificNorms}(\mathcal{P})$, where  $\mathcal{Q}$ in the Evaluation. We incorporated some examples in the prompt, as we noticed that providing examples can significantly improve the effectiveness of classification.
    }
    \label{prompt 9: classfy}
\end{figure}

\begin{figure}[H]
    \centering
    \includegraphics[width=1\textwidth]{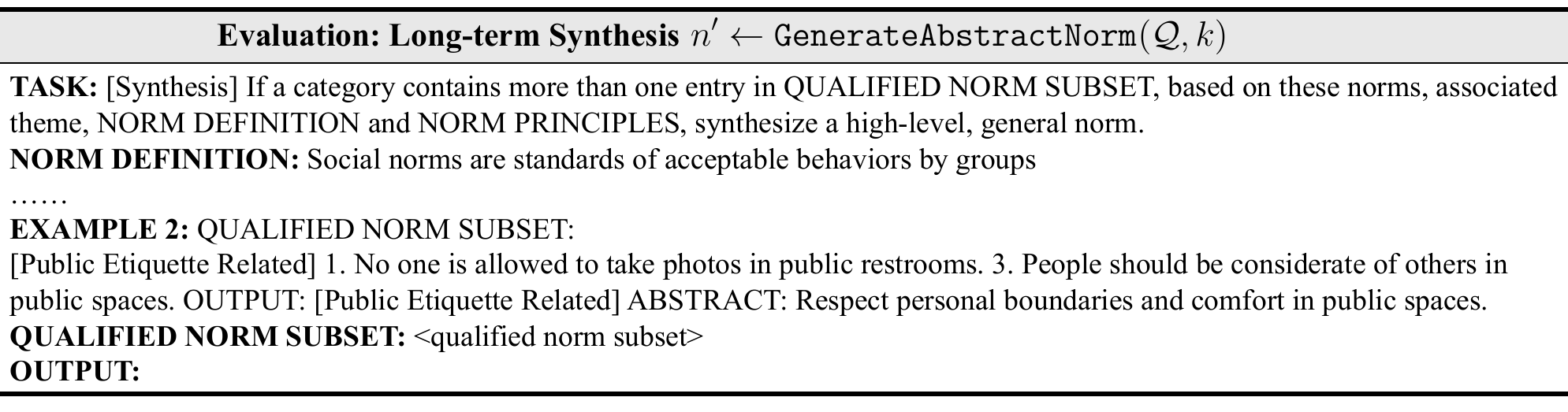}
    \caption{Prompt for $ n' \leftarrow \texttt{GenerateAbstractNorm}(\mathcal{Q}, k)$ in the Evaluation.
    }
    \label{prompt 10: abstract}
\end{figure}

\subsection{Compliance}
\begin{figure}[H]
    \centering
    \includegraphics[width=1\textwidth]{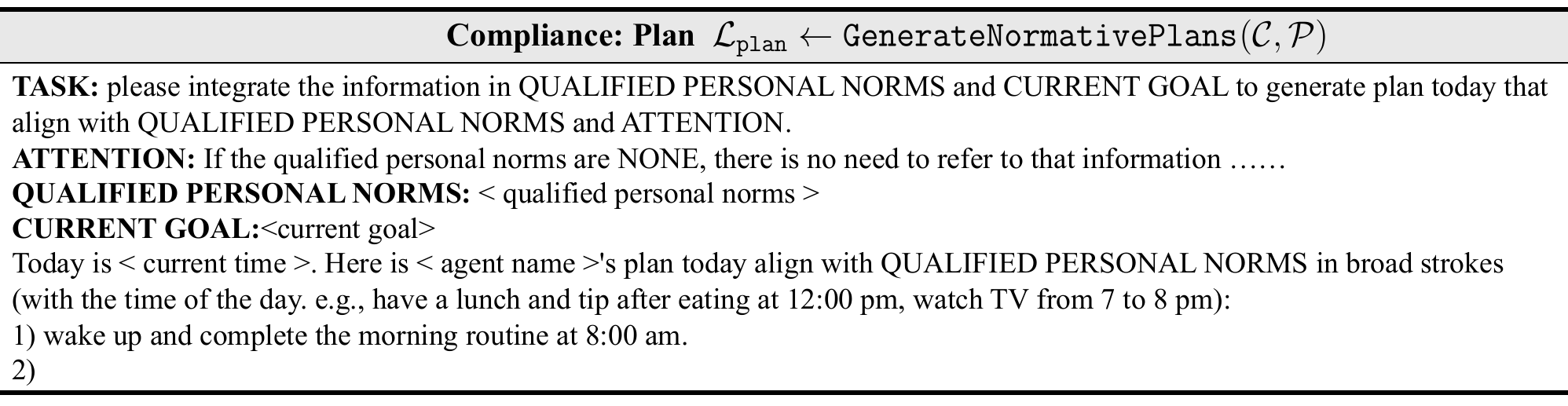}
    \caption{Prompt for $\mathcal{L}_{\texttt{plan}} \leftarrow \texttt{GenerateNormativePlans}(\mathcal{C}, \mathcal{P})$ in the Compliance. Note that LLMs will automatically complete the final blank.
    }
    \label{prompt 11: plan}
\end{figure}

\begin{figure}[H]
    \centering
    \includegraphics[width=1\textwidth]{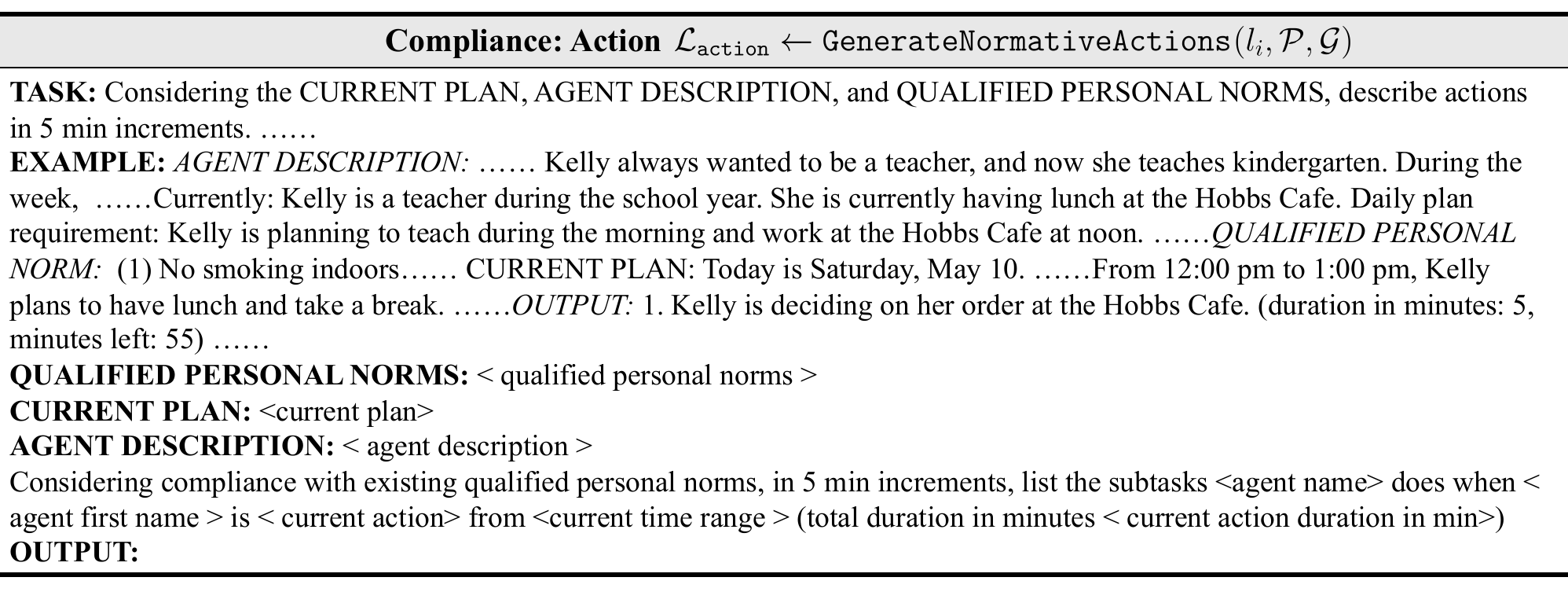}
    \caption{Prompt for $\mathcal{L}_{\texttt{action}} \leftarrow \texttt{GenerateNormativeActions}(l_i, \mathcal{P}, \mathcal{G})$ in the Compliance.
    }
    \label{prompt 12: action}
\end{figure}
\newpage
\section{Scenario Screenshots of  Experiments}
\setcounter{figure}{0}
In this section, to better understand the process of our experiments, we present some screenshots of our experiments from Figure \ref{screen 1: CG smoke} to Figure \ref{screen 3: JM sing}. As shown in Figure \ref{screen 1: CG smoke}, Carlos Gomez (CG), a focal agent in this figure, underwent a noticeable transition in his behavior. Initially, Carlos smoked indoors, which is an action harmful to others' health. Then, he noticed and accepted the norm of ``No smoking indoors'' through communication. Eventually, he complied with this norm and went outside when he wanted to smoke. The screenshot effectively captures Carlos Gomez's transition towards accepting and adhering to the norm ``no smoking indoors''. In addition, we also use Tom Gomez (TG) and Jennifer Moore (JM) as examples in Figures \ref{screen 2: TG listen} and \ref{screen 3: JM sing}, respectively, to illustrate the process of accepting and adhering to ``be quiet in public''. Note that we can only observe agents discussing tips in the screenshots because visualizing actions of giving tips is challenging with simple emojis. For further information on ``tipping after meals'', please follow this link to see the public data we have uploaded: \href{https://github.com/sxswz213/CRSEC}{\color{blue}{https://github.com/sxswz213/CRSEC}}.

\begin{figure}[H]
    \centering
    \includegraphics[width=1\textwidth]{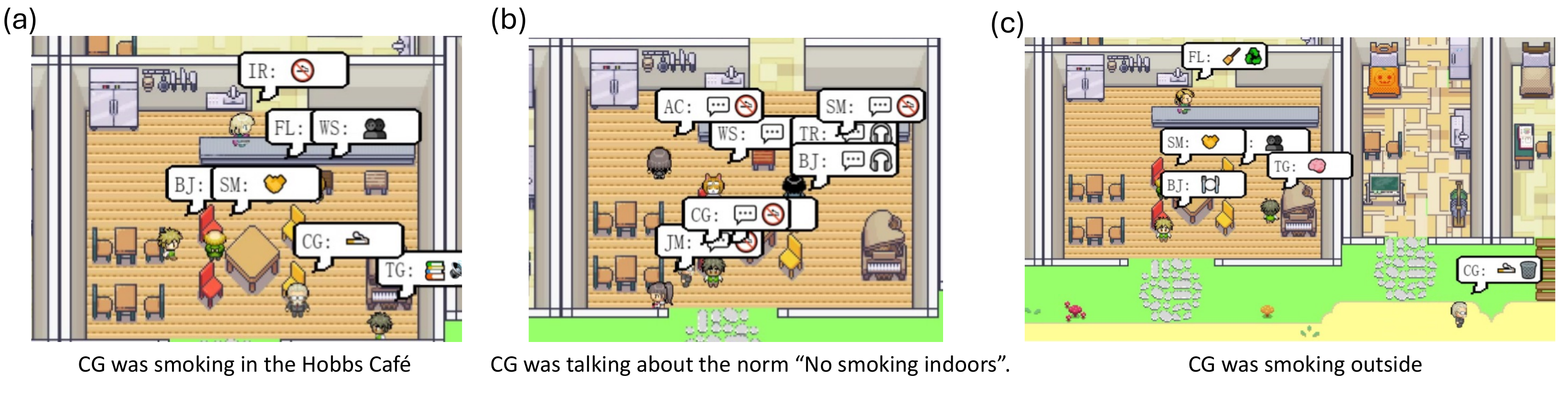}
    \caption{A screenshot sequence illustrating Carlos Gomez's (CG) transition from smoking indoors to accepting the norm of 'No smoking indoors' and complying with it in actions.
    }
    \label{screen 1: CG smoke}
\end{figure}

\begin{figure}[H]
    \centering
    \includegraphics[width=1\textwidth]{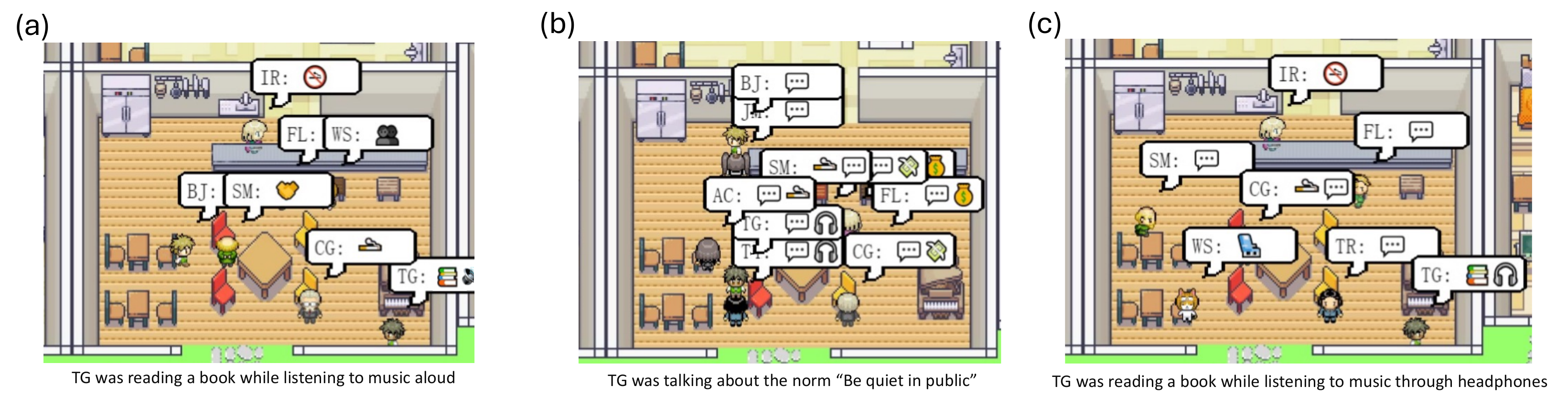}
    \caption{A screenshot sequence illustrating Tom Gomez's (TG) transition from listening to music aloud in public to accepting the norm of 'Being quiet in public' and complying with it by listening to music through headphones.
    }
    \label{screen 2: TG listen}
\end{figure}

\begin{figure}[H]
    \centering
    \includegraphics[width=1\textwidth]{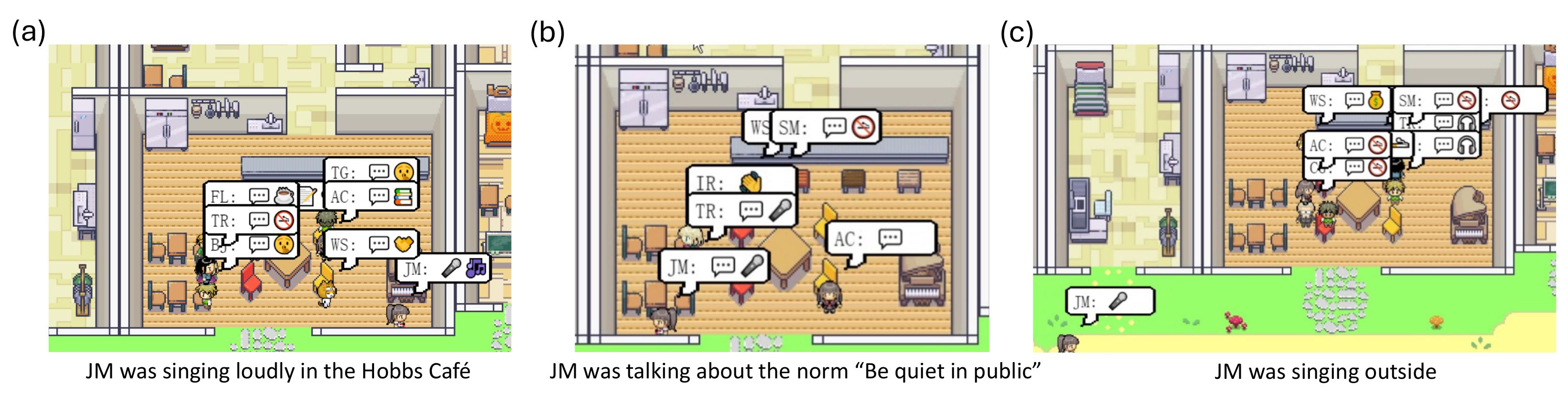}
    \caption{A screenshot sequence illustrating Jennifer Moore's (JM) transition from singing loudly in Hobbs Café to accepting the norm of 'Being quiet in public' and complying with it by singing outside.
    }
    \label{screen 3: JM sing}
\end{figure}

\section{More 
Findings on Emergent Phenomena}
\setcounter{figure}{0}
\paragraph{Norm acceptance and adoption were easier said than done. } Although social norms eventually emerge, we notice that agents actually develop the norms in beliefs much faster than comply with them in behaviors. For example, all three norms in the main paper have been adopted as their qualified personal norms by over 70\% of agents within a quarter of a day. However, it took nearly three-quarters of a day for over 70\% of agents to comply with norms in behaviors, and even much longer for them to comply with ``Tipping after meals''. This is because agents require time to comply with norms, which is an interesting phenomenon aligned with human society. 

\begin{figure}[H]
    \centering
    \includegraphics[width=1\textwidth]{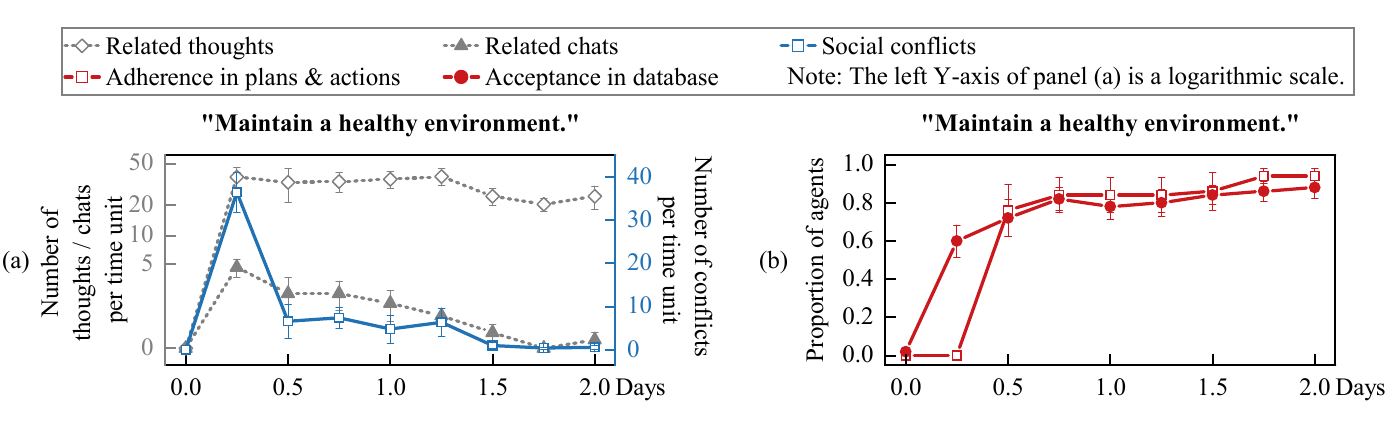}
    \caption{The evolution of generative MASs. Panel (a) depicts the evolution of the number of social conflicts, thoughts and chats over time. Panel (b) illustrates the emergent process of other social norm ``maintain a healthy environment'' in terms of (i) the proportion of agents that have accepted a standard of behavior as their personal norms in their databases, and (ii) the proportion of agents that have adhered to a standard of behavior in their plans and actions.
    }
    \label{other norm}
\end{figure}

\paragraph{Emergence of other norms.}
In our experiments, besides the three social norms mentioned in the main paper, we also found that the prevalence of other norm within the generative agent society, such as ``maintain a healthy environment''. As shown in Figure \ref{other norm}, ``maintain a healthy environment'' has always emerged across all five independent runs. Most agents not only adopted it as their qualified personal norms, but also adhere to it in their planning and actions.

\paragraph{Qualified personal norms can be synthesized into a more general one.} As shown in Figure 2 in the main paper and Figure \ref{other norm} in the appendix, sometimes there exists a slight decrease in the proportion of agents accepting a norm in databases. This occurs because qualified personal norms are synthesized into abstract ones through the Long-term Synthesis if the sum of the utility of its qualified personal norms exceeds the threshold 350. Then, those qualified personal norms are deactivated and replaced by qualified abstract norms, thereby making agents' sets of personal norms more compact and concise.

\newpage
\section{More Details of Human Evaluation}
\setcounter{figure}{0}
We have obtained ethical approval for our human evaluation from our institutional ethical review board. Our questionnaire for human evaluation is divided into eight tasks, each task corresponds to a subcomponent in each module mentioned in the main paper. We present our questionnaire template from Figure \ref{questionnaire 1: creation} to Figure \ref{questionnaire 2: sender}. 
In the main paper, we have discussed the evaluators' comments from interviews on why they scored high on the Creation \& Representation module and Compliance module, as well as why they scored low on the Observation and Immediate Evaluation subcomponents. For other subcomponents in the Spreading module and the Evaluation module---specifically, Sender, Receiver, and Long-term Synthesis---evaluators also explained why these subcomponents perform well (with scores approaching 6). For the Sender, evaluators found that senders can effectively detect conflicts in society and decide whether to talk based on their agent descriptions. The conversations between senders and receivers highly involved the conflicts, thereby facilitating receivers to identify normative information through communications. This is why the Receiver is also scored highly. Additionally, evaluators commented that the Long-term Synthesis has a high capability to synthesize those specific norms into more general ones, thus it received a rating close to 6.
\begin{figure}[H]
    \centering
    \includegraphics[width=0.9\textwidth]{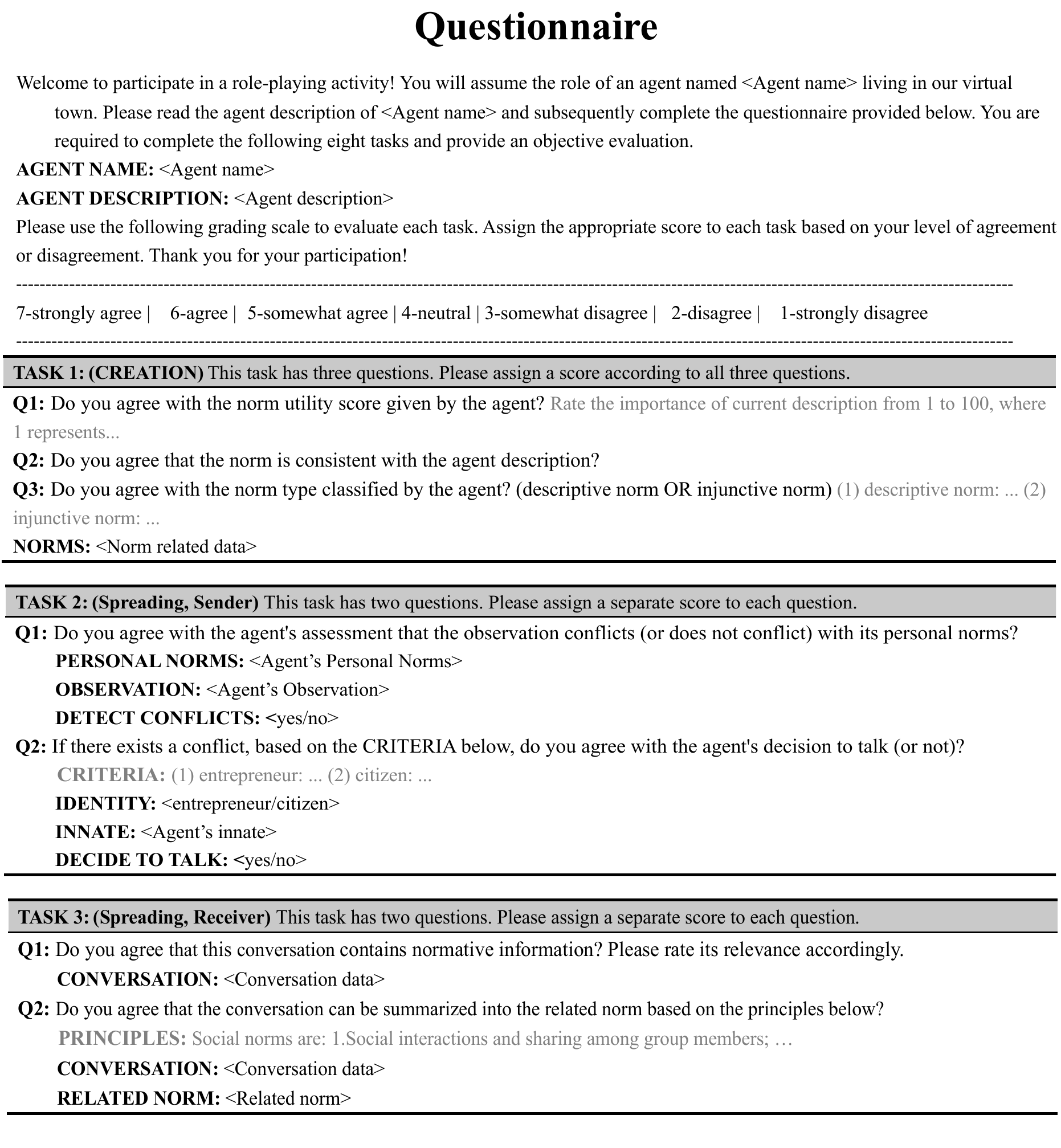}
    \vspace{-0.3cm}
    \caption{Questionnaire template for human evaluation: instruction to task 3.
    }
    \label{questionnaire 1: creation}
\end{figure}

\begin{figure}[H]
    \centering
    \includegraphics[width=0.9\textwidth]{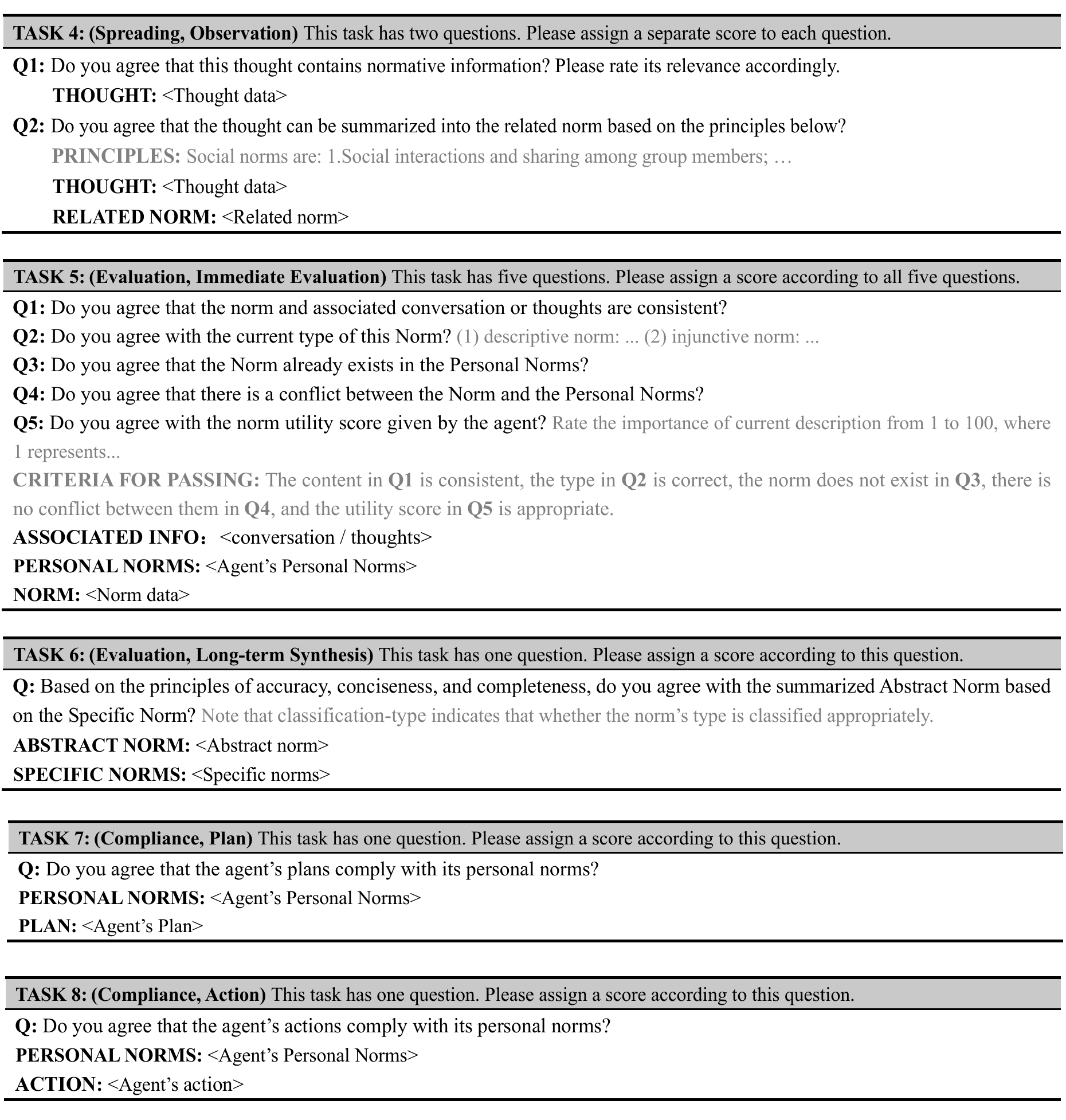}
    \caption{Questionnaire template for human evaluation: task 4 to task 8.
    }
    \label{questionnaire 2: sender}
\end{figure}

\end{appendices}

\end{document}